\documentclass[twocolumn,english,aps,prx,floatfix,amssymb,superscriptaddress,longbibliography]{revtex4-1}
\usepackage[latin9]{inputenc}
\usepackage{verbatim}
\usepackage{float}
\usepackage{amsmath}
\usepackage{amssymb}
\usepackage{graphicx}
\usepackage{color}
\usepackage{xcolor}
\usepackage{tikz}
\newcommand{\ket}[1]{\ensuremath{\left| #1 \right>}}

\newcommand{\Tr}{\text{Tr}}
\usepackage{bbold}

\usepackage[bookmarks=false,linkcolor=blue,urlcolor=blue,colorlinks,citecolor=blue]{hyperref}
\makeatletter

\newcommand{\be}{\begin{equation}}
\newcommand{\ee}{\end{equation}}
\newcommand{\bea}{\begin{eqnarray}}
\newcommand{\eea}{\end{eqnarray}}

\begin{document}

\title{Emergent boundary supersymmetry in a one dimensional superconductor}
\author{Parameshwar R. Pasnoori}
\affiliation{Department of Physics, University of Maryland, College Park, MD 20742, United
States of America}
\affiliation{Laboratory for Physical Sciences, 8050 Greenmead Dr, College Park, MD 20740,
United States of America}
\author{Patrick Azaria}
\affiliation{Laboratoire de Physique Th\'eorique de la Mati\`ere Condens\'ee, Sorbonne Universit\'e and CNRS, 4 Place Jussieu, 75252 Paris, France}
\author{Colin Rylands}
\affiliation{SISSA and INFN Sezione di Trieste, via Bonomea 265, 34136  Trieste, Italy}
\affiliation{Centre for Fluid and Complex Systems, Coventry University, Coventry, CV1 2TT, United Kingdom}
\author{Natan Andrei}
\affiliation{Department of Physics and Astronomy, Rutgers University, Piscataway, NJ 08854-8019, United States of America}

\begin{abstract}
The interplay between bulk  properties and boundary conditions in  one-dimensional quantum systems, gives rise to many intriguing phenomena. These include the emergence of zero energy modes which are of significant interest to a variety of fields. In this work we investigate the presence of such zero modes in cases where the boundary conditions are dynamical and arise due to the coupling to some quantum degrees of freedom. In particular, we study a one-dimensional spin-singlet superconductor, modeled by the Gross-Neveu field theory, coupled to spin $\frac{1}{2}$ magnetic impurities at its boundaries via a spin-exchange interaction.  We solve the model exactly for arbitrary values of the bulk and the impurity coupling strengths using nested coordinate Bethe ansatz and show that the system exhibits a rich boundary phase structure. For a range of couplings, the low energy degrees of freedom form irreducible representations of the supersymmetric $spl(2,1)\otimes spl(2,1)$ algebra which become degenerate at a specific point, indicating the emergence of supersymmetry in the low energy boundary degrees of freedom. We show that at the supersymmetric point there exist exact zero energy modes that map one ground state with the other. We express these in terms of the generators of the algebra.
\end{abstract}
\maketitle

\section{Introduction}
\label{sec:intro}
Zero energy modes (ZEM) have received a lot of attention in the past decade. This is partly due to a crucial role they play in characterizing different phases of matter \cite{wen,AKLT,HALDANE1983,Keselman2015}. In addition, they also have potential applications in quantum computation \cite{NayakMZM,Alicea_2012,pasnoori2025circuitSG}. In one dimension, they arise in systems that exhibit a symmetry protected topological phase (SPT) \cite{Keselman,Boulat,chen2017flux,Diehl2015,Starykh2000,Ruhman2012,Zoller2013,Starykh2007,nijs,pollman,pasnoori2025duality} and also in systems that exhibit spontaneous symmetry breaking (SSB) of a discrete symmetry \cite{Fendley}. The most prominent example of an SPT phase is the Haldane chain \cite{HALDANE1983,AKLT}, which hosts free spin $\pm 1/2$ edge modes at its boundaries giving rise to a four fold degenerate ground state. One of the most famous examples of an SSB phase is the anti-ferromagnetic spin $1/2$ XXZ chain in the gapped phase, where the $\mathbb{Z}_2$ spin flip symmetry is spontaneously broken, giving rise to two fold degenerate ground state \cite{xxzbound2019}. Recently, these ground states were shown to host spin $\pm 1/4$ at the boundaries \cite{XXZpaper}. In both systems mentioned above, the boundaries are free, and the phases exhibited by the system are dictated purely by the type of interactions in the bulk.

\begin{center}
\begin{figure}[!h]
\includegraphics[width=1\columnwidth]{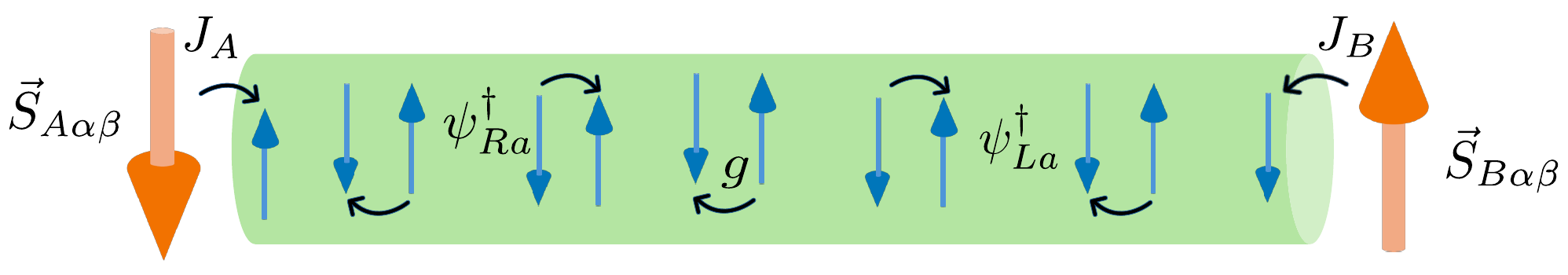}
\caption{Superconducting wire (green) is coupled to a magnetic impurity (orange) at each edge. The electrons in the superconductor interact with each other with strength $'g'$, and interact with the impurities at the left and right edges with strengths $J_A$ and $J_B$ respectively. }
\label{fig:picture}
\end{figure}
\end{center}

As a complement, one can also investigate how different boundary conditions effect the types of phases which can be supported in a particular bulk system.
For example, in certain one-dimensional superconductors, the interplay between the bulk interactions and boundary conditions plays a crucial role. When the boundaries are free, the spin singlet superconductor exhibits a trivial phase, with a unique ground state. In contrast, it exhibits a topological phase when strong magnetic fields are applied at both boundaries.  Its topological nature is unveiled via a duality mapping to a topological, spin triplet superconductor with free boundaries \cite{PAA2}. In this topological phase, the system gives rise to a number of ZEMs, resulting in a four-fold degenerate ground state. In addition, each ground state is shown to exhibit fractionalized spin $\pm 1/4$ at each boundary \cite{Keselman2015,PAA1}. 

The boundary conditions mentioned above are non-dynamical and classical, meaning that they are strictly imposed microscopically and do not fluctuate.   One could, however, investigate scenarios in which the boundaries are dynamical and fluctuate in response to the bulk properties.  A natural setting for this, is to couple the edges of the bulk to quantum mechanical systems with their own internal degrees of freedom. Interesting and fundamental phenomena arise in such systems even if the bulk is non-interacting, as is the case for the famous Kondo effect~\cite{JunKondo}, where bulk fermions compete to screen a magnetic impurity resulting in a strongly correlated many-body ground state.   In the presence of bulk interactions and dephasing, new phases can emerge and have received a lot of attention in the recent past studied \cite{zarand1,KondoXXX,kattel2024,NHK2,kattel2024spin,zarand2}. 

In this work we investigate the role of quantum fluctuating boundary conditions in one-dimensional superconductors and ask whether a spin singlet superconductor exhibits ZEMs in the presence of such boundaries instead of classical magnetic fields mentioned above. In particular, we study a model consisting of a charge  conserving, spin singlet one-dimensional superconductor which is coupled to spin $1/2$ magnetic impurities at its boundaries via a spin-exchange interaction.  We solve this model using nested coordinate Bethe Ansatz for all values of the couplings and uncover a rich structure of different phases arising from the interplay between bulk and boundary interactions and which may exhibit ZEMs as well as finite energy boundary states.  

Aside from the bulk superconducting gap, $\Delta$,  we find that there exist two renormalization group (RG) invariant parameters $d_{L,R}$ corresponding to the left and the right boundaries, which are functions of the bulk and the respective boundary coupling strengths. Depending on the values of $d_{L,R}$, the system exhibits different phases as shown in Figure \ref{fig:qualitative}. The RG invariant parameters $d_m, m=L,R$ can take real values $d_m\in [0,\infty)$ and also imaginary values $d_m=ia_m$, $a_m\in [0,\infty)$. For all real values of $d_m$ and imaginary values $0\leq a_m\le 1/2$, the respective impurity is in the \textit{Kondo phase}, where it is screened by the Kondo effect \cite{Andrei80,AndreiRMP}. This is characterized by the emergence of a \textit{renormalized} Kondo temperature $T_{m}^K\geq \Delta$ (see Eq. (\ref{tkdef})). As the respective impurity coupling strength is decreased, $T_{m}^K$ decreases towards its minimum value $T_{m}^K=\Delta$, which occurs when $a_m=1/2$, where a smooth cross-over between many-body screening to a single particle screening of the impurity takes place. The respective impurity is then said to be in a \textit{Yu-Shiba-Rusinov (YSR) phase} \cite{Yu,Shiba,Rusinov,YSRKondo}, which persists for $a_m\in (1/2,3/2)$. In this phase, the impurity is screened, not by  the many body Kondo cloud, but rather, by a bound state localized near the impurity. This bound state can be removed, and hence the impurity can be unscreened via a single particle excitation in contrast to the Kondo phase. This bound state carries a spin $1/2$ and has energy $E^{m}_B=-\Delta \sin(a_m\pi)$. Hence, for $a_m\in (1/2,1)$, the state in which the impurity is screened has lower energy whereas, the state in which the impurity is unscreened has higher energy. When $a_m\in (1,3/2)$, state in which the impurity is unscreened has lower energy whereas the state in which the impurity is screened has higher energy. When $a_m=1$, these two states are degenerate and the impurity undergoes a first order phase transition between the screened and the unscreened ground states. Note that in the YSR phase, there exist a total of three states corresponding to each respective boundary: one in which the impurity is screened and two in which the impurity is unscreened with spin of the impurity pointing in the upward or downward $z$- direction. These three states form a representation of the supersymmetric $spl(2,1)$ algebra. 
 When both the impurities are in their respective YSR phases, the boundary Hilbert space consists of nine states which form irreducible representations of $spl(2,1)\otimes spl(2,1)$ algebra. At the point $a_A=a_B=1$ these nine states are degenerate and the system possesses six ZEMs. All states are degenerate and accordingly the system is supersymmetric.   As the impurity coupling strength is decreased further such that $a_m>3/2$, the respective impurity enters its \textit{unscreened phase}, where the impurity cannot be completely screened.  Here, the bound states are absent and the supersymmetric structure disappears \footnote{In the unscreened phase, there exists residual antiferromagnetic coupling between the impurity and the bulk, but the interaction strength is not strong enough to completely screen the impurity. We do naturally expect that the impurity is partially screened, but eventually becomes unscreened deep in the unscreened phase $a\gg1$.}. 

We derive and elucidate these results in the remainder of the paper, which is organized as follows. In section~\ref{sec:Hamiltonian} we introduce our model and discuss some of its properties.  In section (\ref{sec:Betheansatz}) we describe the construction of the Bethe ansatz wavefunction for the Hamiltonian~\eqref{Hamiltonian}, and present the resulting Bethe ansatz  equations which encode its spectrum. In section (\ref{sec:results}), we present the phase diagram and discuss the results. In section (\ref{sec:susy}), we discuss the emergent supersymmetric algebra associated with the low energy, boundary Hilbert space. In section (\ref{sec:conclusion}) we conclude and discuss open questions. Some technical details on both the Bethe ansatz construction and the supersymmetric algebra are presented in the appendices.

\section{Hamiltonian}
\label{sec:Hamiltonian}
The one dimensional spin singlet superconductor is described by the Gross-Neveu model \cite{AndreiLowensein81}, whose Hamiltonian is given by $H_{\text{GN}}\!\!=\!\!\int_{-L/2}^{L/2}\mathrm{d}x \mathcal{H}_{\rm GN} (x)$
\bea\nonumber
\mathcal{H}_{\text{GN}}(x)\!\!&=& \sum_{a=\uparrow,\downarrow}\psi^{\dagger}_{L a}(x)i\partial_x \psi^{}_{L a} (x)-\psi^{\dagger}_{R a}(x)i\partial_x \psi^{}_{R a}(x) \\\nonumber
&&-2g\!\!\sum_{a,b,c,d=\uparrow,\downarrow}\psi^{\dagger}_{Ra}(x)\psi^{\dagger}_{Lc}(x)\psi_{Rb}(x)\psi_{Ld}(x)
\\\label{bulk}
&&\quad\times \left(\sigma^x_{ab}\sigma^x_{cd}+\sigma^y_{ab}\sigma^y_{cd}+\sigma^z_{ab}\sigma^z_{cd}\right),
\eea 
where the fields $\psi_{L(R) a}(x)$, $a=(\uparrow, \downarrow)$, 
describe left and right moving, linearly dispersing fermions carrying spin $1/2$. We set their velocity to $1$. The fermions in the superconductor interact with each other through a spin-exchange interaction strength $g>0$. They interact with the spin $1/2$ impurities at the left and the right boundaries through spin-exchange interaction with coupling strengths $J_A$ and $J_B$ respectively. This is described by the Hamiltonian
\bea
\nonumber
H_{\text{imp}}= -J_A\vec{\sigma}_{ab}\cdot\vec{S}_{A\alpha\beta}\psi^{\dagger}_{La}(-L/2)\psi^{}_{Rb}(-L/2)\\-J_B\vec{\sigma}_{ab}\cdot\vec{S}_{B\alpha\beta}\psi^{\dagger}_{La}(L/2)\psi^{}_{Rb}(L/2). \label{kondo}\eea
In the above equations, $\vec{\sigma}_{ab}, \vec{S}_{A\alpha\beta},\vec{S}_{B\alpha\beta}$ are the Pauli matrices acting in the spin spaces of the bulk fermions and spin spaces of the impurity $A$ and $B$ respectively.

The Hamiltonian of the complete system is then described by
\be
\label{Hamiltonian}
H= H_{\text{GN} }+ H_{\text{imp}}.
\ee
In the geometry we are considering, the fermion fields $\psi_{L,R}(x)$ satisfy open boundary conditions (OBC) at both left and right edges which read, 
\be
 \psi_{R}(-L/2)= - \psi_{L}( -L/2), \;  \psi_{R}(L/2)= - \psi_{L}( L/2),
\label{OBC}
\ee
where we have introduced the spinor notation $\psi_{L(R)}^T \equiv (\psi_{L(R) \uparrow}, \psi_{L(R) \downarrow})$. The Hamiltonian (\ref{bulk}) has been solved using Bethe ansatz \cite{AndreiLowenstein79,AndreiLowensein81,DestriLowenstein}. %\colin{\textcolor{brown}{Due to the fermion-fermion interaction, the system exhibts a separation of spin and charge degrees of freedom.. The charge excitations called holons, remain gapless whereas the spin excitations, spinons, acquire a mass gap, $\Delta$,}
The system exhibits a separation of spin and charge degrees of freedom. Moreover, due to the fermion-fermion interaction, the system exhibits a mass gap $\Delta$ in the spin sector where the excitations are called spinons. The charge sector however remains gapless and the charge excitations are called holons. The mass gap $\Delta$ in the Bethe Ansatz cutoff scheme reads,
\bea
\Delta&=&D\arctan{\big([\sinh{\pi b}]^{-1}\big)}, \\
b&=&(1-g^2/4)/2g,
\eea
where $D$ is a cut-off scale. The mass gap stabilizes quasi-long range spin-singlet superconducting order, characterized by 
\bea \langle {\cal O}_s(x) {\cal O}_s(0) \rangle \sim |x|^{-1/2}, \\  {\cal O}_s \propto \psi^{\dagger}_{R\uparrow}\psi^{\dagger}_{L\downarrow} - \psi^{\dagger}_{R\downarrow}\psi^{\dagger}_{L\uparrow}.\eea 
The case of one impurity is also integrable for arbitrary bulk and impurity coupling strengths, and has been studied in \cite{PRA,YSRKondo,zarand1,zarand2}. These exhaustive studies have focused on the nature of the impurity physics, where it was shown that due to the competition between the impurity and superconductivity, the system exhibits different phases which are characterized by the nature of the screening of the impurity or lack thereof. In this work we expand upon these previous studies and  solve the model (\ref{Hamiltonian}) for arbitrary coupling strengths $g, J_A$ and $J_B$ using Bethe ansatz and show that the system exhibits a more complex phase structure mentioned above including emergence of boundary supersymmetry.

\section{Bethe anstaz}
\label{sec:Betheansatz}
We construct the eigenstates of~\eqref{Hamiltonian} using the standard coordinate Bethe ansatz approach which relies upon the existence of a conserved $U(1)$ charge. 
In the present case, the Hamiltonian has fermion number symmetry which is generated by the total fermion number operator $\hat N$,
\be
\label{N}
\hat{N}=  \int_{-L/2}^{L/2} dx\;  (\psi^{\dagger}_L(x) \psi^{}_L(x) + \psi^{\dagger}_R(x) \psi^{}_R(x)). 
\ee
In addition, the  system is $SU(2)$ invariant, as it commutes with the total spin operator 
\be
\vec S_{\rm T}= \vec s + \vec S_{},
\ee
where $\vec S $ is the impurity spin operator and $\vec s_{} =  \int_{-L/2}^{L/2} dx\; \vec s_{}(x)$ with
\be
\label{S}
\vec s_{}(x)= \frac{1}{2} (\psi^{\dagger}_L(x) \vec \sigma \psi^{}_L(x) + \psi^{\dagger}_R(x) \vec \sigma \psi^{}_R(x)),
\ee
that of the bulk fermions. Since the Hamiltonian commutes with total particle number $\hat N$, we can diagonalize $H$ by constructing the exact eigenstates in each  $N$-particle sector. The $N$-particle eigenstate whose energy is given by $E=\sum_{j=1}^Nk_j$, where $k_j$ are momenta of the particles, takes the form of superpositions of plane waves for different orderings of the particles. In general, it is expressed as,
\begin{eqnarray}\nonumber
\ket{\{k_j\}}=
\sum_{Q,\vec{a},\vec{\sigma}}\int \theta(x_Q) A^{\{\sigma\}}_{\{a\}}[Q] \prod_j^{N_e} e^{i\sigma_j k_jx_j}\\\psi^{\dagger}_{a_j\sigma_j}(x_j)\ket{0}
 \otimes \ket{a_A}\otimes \ket{a_B}.\label{NparticleS}\end{eqnarray}
Here, we sum over all  spin and chirality configurations specified by $\{a\}=\{a_1,\dots ,a_N,a_A,a_B\}$, $\{\sigma\}=\{\sigma_1,\dots ,\sigma_N\}$ as well as different different orderings of the $N$ particles. These different orderings correspond to elements of the symmetric group $Q\in \mathcal{S}_N$. In addition, $\theta(x_Q)$ is the Heaviside function which is non-zero only for that particular ordering. The amplitudes $A^{\{\sigma\}}_{\{a\}}[Q]$ are related to each other by the various $S$-matrices. These $S$-matrices are derived in the appendix, and are given by
\begin{eqnarray}\label{S12}S^{ij}= \frac{2ib \; I^{ij} +P^{ij}}{2ib+1},~
S^{jm}=\frac{\; I^{jm} -i c_m P^{jm}}{1-ic_m}.\end{eqnarray}
Where $I^{ij}$ is the identity operator
 and $P^{ij}=(I^{ij}+ \vec{\sigma}_i\cdot\vec{\sigma}_j)/2$ is the permutation operator acting on the spin spaces of particles $i$ and $j$ with $m=A,B$ indicating the impurity $A$ and $B$ respectively. Here the parameters $c_m$ are  related to the impurity coupling strengths $J_m$ as $c_{m}=\frac{2J_{m}}{1-3J_{m}^2/4},~m=A,B$. $S^{jm}$ is the $S$-matrix between particle $j$ and the impurity $m=A,B$, describing the interaction of the electrons with the impurities. It relates amplitudes which  differ by   changing the chirality of the right most particle $+\to -$ or the leftmost quasiparticle $- \to +$. Similarly, amplitudes which are related by swapping the order of particles with different chiralities are related by the particle-particle $S$-matrix, $S^{ij}$. An additional  $S$-matrix, denoted by $W^{ij}$, is also required. It  relates amplitudes that differ by exchanging particles of the same chirality. This is given by $W^{ij}=P^{ij}$. The consistency of the solution is then guaranteed as the $S$-matrices satisfy the Yang-Baxter and reflection equations~\cite{Sklyannin, Cherednik},
\bea\label{YB1}
W^{jk} \;W^{ik}\; W^{ij} &=& W^{ij} \;W^{ik} \;W^{jk},\\ \label{YB2}
S^{jk}\;S^{ik}\;W^{ij} &=& W^{ij}\;S^{ik}\;S^{jk},
\\ S^{jB}\;S^{ij}\;S^{iB}\;W^{ij}&=&W^{ij}\;S^{iB}\;S^{ij}\;S^{jB}\label{YB3},\\S^{jA}\;S^{ij}\;S^{iA}\;W^{ij}&=&W^{ij}\;S^{iA}\;S^{ij}\;S^{jA}\label{YB4}.\eea
 Here, as perviously, the superscripts denote which particles the operators act upon. 
 
 Imposing the boundary condition at $x=-L/2$ quantizes the single particle momenta $k_j$ and allows us to determine the spectrum of $H$. Relegating the details to appendices, we have 
\begin{eqnarray}\label{energy}
e^{-ik_jL}\!=\!\prod_{\alpha=1}^Mf(2b, 2\lambda_\alpha),~
f(x,z)=\!\prod_{\sigma=\pm}\frac{x+\sigma z+i}{x+\sigma z-i}
\end{eqnarray}
where $M\leq N/2$ and the parameters,  $\lambda_\alpha$, $\alpha=1,\dots,M$ are known as Bethe roots and  satisfy the Bethe Ansatz equations
\begin{eqnarray}\label{BAE}
\left[f(2\lambda_\alpha,2b)\right]^{N}\prod_{m=A,B}f(2\lambda_{\alpha},2d_m)=\prod_{\alpha\neq \beta }^Mf(\lambda_\alpha,\lambda_\beta),
\end{eqnarray}
with,
\begin{align} \label{RGparam}d_m&= \sqrt{b^2-2b/c_m-1}~.\end{align} 
 The Bethe roots govern the spin degrees of freedom of the system,   and $M\leq N/2$ gives the total $z$-component of spin, $S^z=N/2-M$. The solutions $\lambda_\alpha$ can be real or take complex values in the form of strings. In order to have a non vanishing wavefunction they must all be distinct, $\lambda_\alpha \neq \lambda_\beta$. In addition, the value $\lambda_\alpha=0$ should also be discarded as it results in a vanishing wavefunction \cite{ODBA}. Bethe equations of the type \eqref{BAE} are reflection symmetric, that is, they are invariant under $\lambda_\alpha\rightarrow -\lambda_\alpha$ transformation. Due to this symmetry, solutions to the Bethe equations occur in pairs $\{-\lambda_\alpha,\lambda_\alpha\}$ \cite{XXXkondo}. Taking the  logarithm of \eqref{energy} and \eqref{BAE} we obtain 
\begin{eqnarray}\label{logEnergy}
k_j=\frac{2\pi n_j}{L}+\frac{2}{L}\sum_{\beta=-M}^M\Theta( b -\lambda_\beta,1/2),
\end{eqnarray}
 and
\begin{align}\nonumber
-\pi I_\alpha&+\!\sum_{\sigma=\pm} N_e\Theta(\lambda_\alpha+\sigma b,1/2)+\!\!\!\sum_{m=A,B}\!\!\Theta(\lambda_\alpha+\sigma d_m ,1/2)\\&+\Theta(\lambda_\alpha,1/2)+ \Theta(\lambda_\alpha,1)\label{Logbae}
=\!\sum_{\beta=-M}^{M} \Theta\left(\lambda_\alpha- \lambda_\beta,1\right).\end{align}
Where, $\Theta(x,z)=\text{arctan}(x/z)$ and $n_j$ and $I_\alpha$ are integers arise from the logarithmic branch and serve as the quantum numbers of the states. The quantum numbers $I_\alpha$ correspond to the spin degrees of freedom while the quantum numbers $n_j$ are associated with the charge degrees of freedom. They obey a Pauli exclusion principle, meaning there can be no repeated spin or charge quantum numbers.  Moreover, $I_\alpha$ and $n_j$ can be chosen independently implying the charge spin decoupling mentioned above.  While the spin quantum numbers are bounded by the form of~\eqref{Logbae}, the charge numbers are not. Therefore, constructing the ground state requires the introduction of a  cutoff such that $\pi|n_j|/L < \pi D$ where $D=N/L$ is the density of fermions \cite{AndreiLowenstein79}. The spectrum of $H$ is found first, with $D$ finite, after which the cutoff is removed by taking $D\to\infty$ while holding some physical scale fixed. The number of coupling constants means that one can remove the cutoff while keeping three scales fixed. These correspond to RG invariants of the model and are the superconducting mass gap $\Delta$ as well as two boundary scales, $d_{A,B}$ given above~\eqref{RGparam}.

We are interested in the attractive regime, $g>0$, and further consider only the case $b>0$. Within this range, the model exhibits several phases depending on the values of $b$, $c_A$ and $c_B$ or equivalently $d_A$ and $d_B$.  $d_m$, $m=A,B$ can be real or purely imaginary, in which case we take $d_m=ia_m$. In the thermodynamic limit, $N,L\rightarrow\infty$, the Bethe roots form a dense set and the Bethe equations can be expressed as integral equations, which can be solved by Fourier transform, for the distributions of the Bethe roots. The solutions to the Bethe equations (\ref{Logbae}) are provided in the appendix. Below we summarize these results.

\section{Boundary Phase diagram}
 \label{sec:results} 
\begin{center}
\begin{figure}[!h]
\includegraphics[width=1\columnwidth]{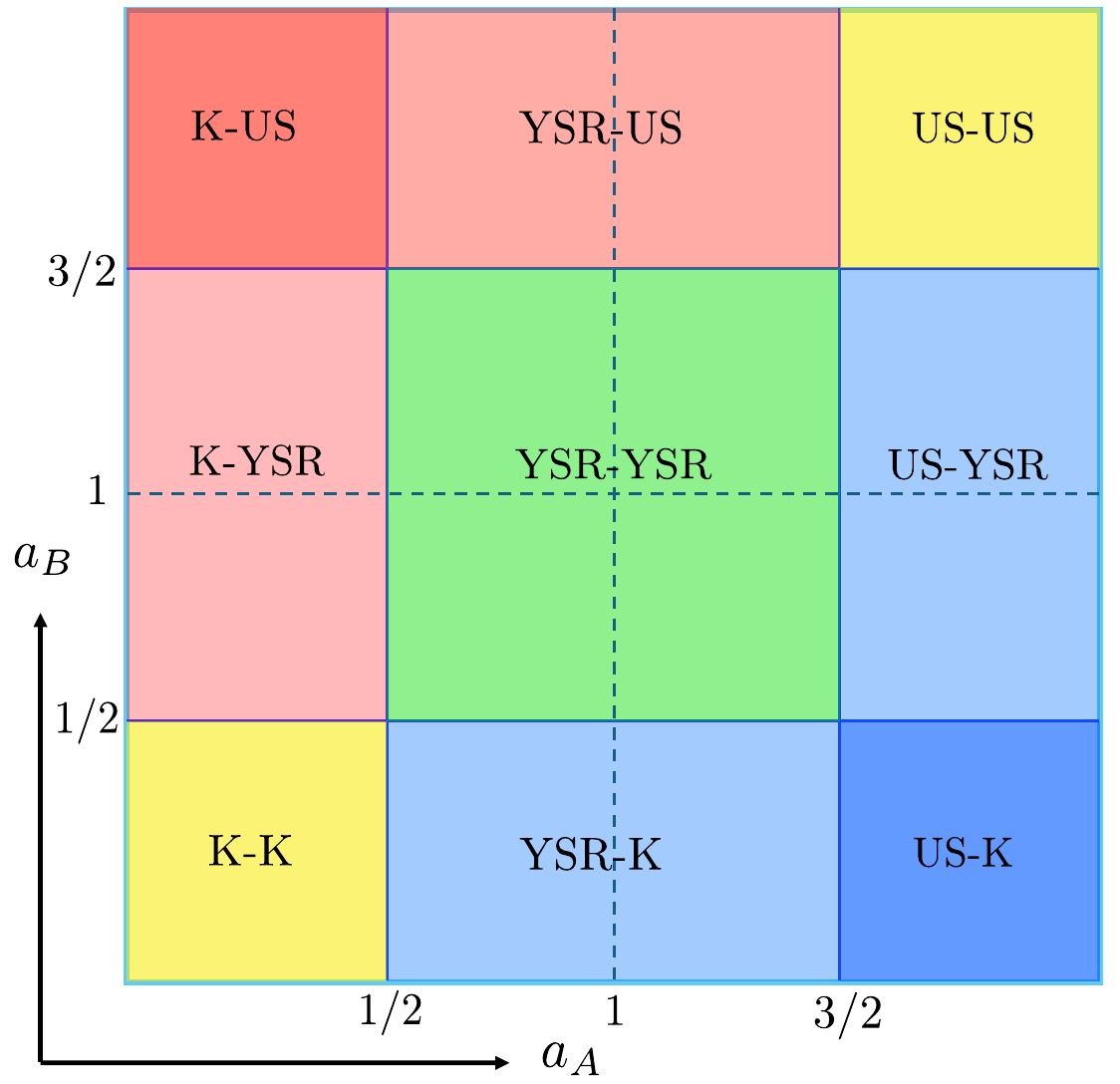}
\caption{ Phase diagram showing nine different boundary phases exhibited by the system. Each phase is labelled by the phase exhibited by the impurity at the left boundary followed by the phase exhibited by the impurity at the right boundary. The x and y axes represent the RG invariant parameters corresponding to the impurities localized at the left and right boundaries respectively. Here `K' stands for Kondo, `US' stands for unscreened and `YSR' stands for Yu-Shiba-Rusinov. The system exhibits mid-gap states at both the boundaries in the YSR-YSR phase. At the supersymmetric point $a_A=a_B=1$, the energy of these mid-gap states is zero, and hence the system exhibits nine-fold degenerate ground state. These nine states form an irreducbible representation of the supersymmetric $spl(2,1)\otimes spl(2,1)$ algebra. }
\label{fig:qualitative}
\end{figure}
\end{center}

   There exist three RG-invariant quantities: The superconducting mass gap $\Delta$ and the parameters $d_A$ and $d_B$. The parameters $d_{A,B}$,  are a function of the respective impurity and the bulk coupling strengths (\ref{RGparam}), and are a measure of the effective coupling strength of the respective impurity with the superconductor. These parameters can take real values $d_{A,B}\in [0,\infty)$, and also imaginary values $d_{A,B}=ia_{A,B},\;  a\in (0,\infty)$. Large real values of $d_{A,B}$ correspond to the respective impurity being very strongly coupled to the superconductor, whereas large imaginary values correspond to weak coupling. The system exhibits several phases which depends on the values of both $d_A$ and $d_B$ as shown in the (Fig: \ref{fig:qualitative}). 

Note that the phases we refer to here pertain to the boundary physics. They are not associated to the breaking of a certain symmetry. Rather, in going from one phase to another there is a complete restructuring of the ground state wave function as well as the Hilbert space of low energy excitations. 

\subsection{Kondo-Kondo (K-K) phase}
When both the parameters take real values in the range $d_{A,B}\in (0,\infty)$ and also imaginary values in the range $a_{A,B}\in [0,1/2)$, the system exhibits the `Kondo-Kondo (K-K) phase'. In the K-K phase, the system exhibits a unique ground state
\be \ket{\rm{K;\!K}} \equiv \ket{0}\otimes \ket{0},
\ee
which is a spin singlet $S_T=0$. We have used the notation where the first ket in the direct product corresponds to the left boundary where the impurity $A$ is located, and the second ket corresponds to the right boundary where the impurity $B$ is located.  We find that the ground state in this phase is  even with respect to the fermionic parity which is given by
\be
\mathcal{P}=(-1)^{\hat{N}},.
\ee
 As the name suggests, each impurity is screened by the Kondo effect, which is characterized by the generation of a strong coupling, energy scale called the Kondo temperature $T^K_{A,B}$ corresponding to the respective impurity. This is defined as the width at half maximum of the ratio of the density of states (DOS) of the respective impurity and that of the bulk per unit length, $R_{A,B}(E)= \frac{L}{2} \frac{ \rho_{\rm imp_{A,B}}(E)}{ \rho_{\rm bulk}(E)}$, where 
\begin{align}\label{DOS}
 R_{A,B}(E)&= \frac{\Delta \cosh{\pi d_{A,B}}}{(E^2+\Delta^2 \sinh^2{\pi d_{A,B}})}, \; E \ge \Delta,\\
 \rho_{\rm bulk}(E)&=\frac{E}{\pi \sqrt{E^2-\Delta^2}}.
 \end{align}
This yields \cite{PRA, YSRKondo}
\be T^K_{A,B}= \Delta \sqrt{1+ \cosh^2{\pi d_{A,B}}}.\label{tkdef}
\ee
When the impurity is strongly coupled, that is when its respective parameter $d_{A,B}\gg 1$, we can infer from the the above equation that  $T^K_{A,B}/\Delta\gg1$. In this regime $R_{A,B}(E)$ takes the characteristic Lorentzian form similar to the case of a metallic bath \cite{AndreiRMP}. As the impurity coupling strength decreases, which corresponds to decreasing the values of the respective parameter $d_{A,B}$, the width of (\ref{DOS}) gets narrower such that $T^K_{A,B}$ decreases. When $d_{A,B}$ is further decreased past the value of $d_{A,B}=0$, where it becomes imaginary $d_{A,B}=ia_{A,B}$, and as $a_{A,B}\rightarrow 1/2$, the respective DOS (\ref{DOS}) becomes sharply peaked at $E=\Delta$. In this limit, the respective Kondo temperature $T^K_{A,B}\rightarrow\Delta$. This can be interpreted as a many-body screening turning into an effective single-particle screening of the impurity i.e. the scale associated to the screening of the impurity is the same as that associated to a single particle excitation. This is in contrast to the deep Kondo phase where $T^K_{A,B}\gg \Delta$, and so cannot be associated to a single excitation.   As the impurity coupling strength is further decreases, the respective impurity undergoes a phase transition into its YSR phase.

\subsection{YSR-YSR phase}
Let us consider the case where both the RG invariant parameters take imaginary values in the range $a_{A,B}\in (1/2,3/2)$. In this range, both the impurities are in their respective YSR phases, where the physics associated with the impurities is drastically different from that of the Kondo phase. At the phase transition point, $a_{A,B}=1/2$, as mentioned above, the impurities are effectively screened by a single particle bound state localized near the respective impurities. When the impurities are in their respective YSR phases, where $a_{A,B} \in (1/2,3/2)$, these bound states are associated with `boundary strings', and we denote this state by $\ket{\rm{S;\!S}}$. This state is a total singlet $S_T=0$ and has even fermionic parity is $\mathcal{P}=+1$. Unlike the K-K phase where the impurity cannot be unscreened by exciting any finite number of particles, in the YSR-YSR phase, the single particle bound states localized near the impurities can be removed, and hence, one can unscreen the impurities. The resultant states $\ket{\rm{S;\!US}}_{\pm 1/2}$, $\ket{\rm{US;\!S}}_{\pm 1/2}$ obtained by removing the bound state at the right and the left edges respectively, in which the impurities B and A respectively are unscreened, have a total spin $S_T=1/2$ and correspond to odd fermionic parity $\mathcal{P}=-1$. Note that each of these states are two fold degenerate corresponding to $S^z=\pm 1/2$ of the unscreened impurity, which is represented by the subscripts. By removing the two bound states at both the edges, one obtains the state $\ket{\rm{US;\!US}}_{\pm 1/2,\pm 1/2}$ in which both the impurities are unscreened. It has even fermionic parity $\mathcal{P}=+1$ and is four-fold degenerate. There exist three states with total spin $S_T=1$ corresponding to the triplet and one state with $S_T=0$ corresponding to the singlet of the two spin $1/2$'s associated with the unscreened impurities. The subscripts denote the spin $S^z$ of the impurity localized at the left and the right boundaries respectively.

These states form a representation of the $spl(2,1)\otimes spl(2,1)$ superalgebra
\be \label{repphy}\ket{\frac{n_{i,A}+n_{b,A}}{2},S_{T,A},S^z_A}\otimes \ket{\frac{n_{i,B}+n_{b,B}}{2},S_{T,B},S^z_B}.
\ee 
Here the first and the second kets correspond to the impurities A and B respectively. $n_{i,A}$, $n_{i,B}$ denote the number of impurities at the left and the right boundaries respectively $(n_{i,A}=n_{i,B}=1)$. $n_{b,A}$ and $n_{b,B}$ denote the number of bound states localized near the impurity $A$ and B respectively. The quantity $\frac{1}{2}(n_{i,A(B)}+n_{b,A(B)})$ is called the baryon number \cite{Rittenberg} corresponding to the left (right) boundaries respectively. $S_{T,A}$ and $S_{T,B}$, denotes the total spin corresponding to the left and the right boundaries respectively and $S^z_A$ and $S^z_B$ denote the respective $z$ component of the spin. In the next section we shall explain in detail the representation theory associated to these states, which are tabulated in Table.\ref{table1}.

\begin{table}[h!]
\centering
\caption{The table shows the all the nine states in the YSR-YSR phase which 
are represented as the direct product of the states associated with each boundary. At each boundary $i=A,B$, there exist three states: One state $\ket{1,0,0}$ in which the impurity is screened by the localized bound state which has total spin (local boundary spin) $S_{T,m}=0$ and two states $\ket{\frac{1}{2},\frac{1}{2},\pm \frac{1}{2}}$ in which the impurity is unscreened which have total spin $S_{T,m}=\frac{1}{2}$.}

\begin{tabular}{|c|c|}
\hline
\hline
  State  & $\ket{\frac{1}{2}(1+n_{b,A}),S_{T,A},S^z_A}$\\ & $\otimes \ket{\frac{1}{2}(1+n_{b,B}){2},S_{T,B},S^z_B}$ \\
\hline
 $\ket{\rm{S;\!S}}$& $\ket{1,0,0}\otimes \ket{1,0,0}$   \\
  $\ket{\rm{S;\!US}}_{\pm 1/2}$ & $\ket{1,0,0}\otimes \ket{\frac{1}{2},\frac{1}{2},\pm \frac{1}{2}}$  \\
$\ket{\rm{US;\!S}}_{\pm 1/2}$ & $ \ket{\frac{1}{2},\frac{1}{2},\pm \frac{1}{2}}\otimes \ket{1,0,0}$  \\
$\ket{\rm{US;\!US}}_{\pm 1/2,\pm 1/2}$& $\ket{\frac{1}{2},\frac{1}{2},\pm \frac{1}{2}}\otimes \ket{\frac{1}{2},\frac{1}{2},\pm \frac{1}{2}}$\\
\hline
\hline
\end{tabular}
\label{table1}
\end{table}

The energy of the bound state localized at the left boundary is equal to the difference of the energy $E_{\ket{\rm{S;S}},N}$ ($E_{\ket{\rm{S;US}}_{\pm 1/2},N}$) of the state $\ket{\rm{S;\!S}}$ ($\ket{\rm{S;\!US}}_{\pm 1/2}$), which has even (odd) number of electrons $N$, and the energy $E_{\ket{\rm{US;S}}_{\pm 1/2},N\pm1}$ ($E_{\ket{\rm{US;US}}_{\pm 1/2,\pm 1/2},N\pm 1}$)  of the state 
$\ket{\rm{US;\!S}}_{\pm 1/2}$ ($\ket{\rm{US;\!US}}_{\pm 1/2,\pm 1/2}$), which has odd (even) number of electrons, $N\pm1$, respectively.
\begin{align}
E_{\text{bound,A}}= E_{\ket{\rm{S};\alpha},N}- \frac{1}{2}\left(E_{\ket{\rm{US};\alpha}, N-1}+E_{\ket{\rm{US};\alpha},N+1}\right),
\end{align}
where $\alpha=\rm {S} \:\text{or}\: \rm{US}$ (We have suppressed the spin indices to ease the notation). The energy $E_{\text{bound,A}}$ of this bound state has the charge and the spin part as mentioned before, where $E_{\text{bound,A}}=E_{\text{charge}}+E_{\text{spin,A}}$, where 
\be\label{Echarge}
E_{\rm charge} = -\frac{\pi}{2L}, \;\; 
E_{\text{spin,A}} = -\Delta \sin{\pi a_A}.
\ee
 We can infer from the above expression that one can remove the bound state localized near the impurity $A$ from the state $\ket{\rm{S;\!S}}$ ($\ket{\rm{S;\!US}}_{\pm 1/2}$) and obtain the state $\ket{\rm{US;\!S}}_{\pm 1/2}$ ($\ket{\rm{US;\!US}}_{\pm 1/2,\pm 1/2}$), thereby unscreen the impurity $A$. For $1<a_A<3/2$, one can add this bound state to the state $\ket{\rm{US;\!US}}_{\pm 1/2}$ ($\ket{\rm{US;\!US}}_{\pm 1/2,\pm 1/2}$) and obtain the state 
 $\ket{\rm{S;\!S}}$ ($\ket{\rm{S;\!US}}_{\pm 1/2}$) thereby screen the impurity $A$. Both of these processes cost energy $E_{{\rm{bound}},A}$, which is always smaller than the mass gap $\Delta$. 

 Similarly, The energy of the bound state corresponding to the right boundary, is given by the (\ref{Echarge}) with $A\rightarrow B$. The ground state of the system in this phase depends on the values of $a_{A,B}$ and is tabulated in table (\ref{table2}). Depending on the values of $a_{A,B}$, the lowest energy excited states on top of the respective ground state are obtained by either adding or removing the bound states localized at the boundaries. Since the energy of the bound states is always less than the mass gap $\Delta$, these are called the mid-gap states. At $a_{A}=1$ and at $a_{B}=1$, the respective impurity undergoes a first order quantum phase transition across which the ground state of the system changes. Notice that in the thermodynamic limit $L\rightarrow \infty$, $E_{\text{charge}}\rightarrow 0$, and hence there exists a genuine zero-energy mode (ZEM) at $a_A=1$ and at $a_B=1$ localized at the respective boundary.

\begin{table}[h!]
\centering
\caption{Table shows the ground state corresponding to the different values of the RG invariant parameters $a_{A}$ and $a_{B}$ in the YSR-YSR phase. At the special values $a_A=1$ or $a_B=1$, the system exhibits a three fold degenerate ground state, where the two states corresponding to the respective impurity being unscreened and the state in which the impurity is screened are degenerate. At the supersymmetric point $a_A=A_B=1$, all the nine states (three states corresponding to each boundary) are degenerate. Note that to ease the notation we have suppressed the spin indices.}

\begin{tabular}{|c|c|c|}
\hline
\hline
  Ground state  & $a_A$ & $a_B$ \\
\hline
 \ket{\rm{S;\!S}} & $(1/2,1)$& $(1/2,1)$\\
\ket{\rm{S;\!US}} & $(1/2,1)$& $(1,3/2)$\\
\ket{\rm{US;\!S}} & $(1,3/2)$& $(1/2,1)$\\
\ket{\rm{US;\!US}} & $(1,3/2)$& $(1,3/2)$\\
\hline
\hline
\end{tabular}
\label{table2}
\end{table}

 \subsubsection{Supersymmetric point}
 Notice that in the YSR-YSR phase, when $a_A=a_B$, the two states $\ket{\rm{S;\!US}}_{\pm 1/2}$ and $\ket{\rm{US;\!S}}_{\pm 1/2}$, in which one of the impurity is screened whereas the other is unscreened belong to the odd fermionic parity sector, and the energy of the bound state corresponding to the left and the right boundaries has the same energy. Due to this, these two states are degenerate up to exponential in the system size $\sim e^{-L}$. These two states are mid-gap states for all values of $a_{A}=a_{B}<1$ and $a_{A}=a_{B}>1$. At the special point where $a_{A}=a_{B}=1$, the energy of the bound state corresponding to both boundaries is zero, and hence these states are degenerate with the states $\ket{\rm{S;\!S}}$ and $\ket{\rm{US;\!US}}_{\pm 1/2,\pm 1/2}$ in which both the impurities are screened and unscreened respectively. Note that, since the states in which both the impurities are either screened or unscreened correspond to even fermionic parity $\mathcal{P}=+1$, the degeneracy is only up to $\sim 1/L$ in the system size. Nevertheless, in the thermodynamic limit, at this point, the states shown in Table (\ref{table1}) are nine fold degenerate. In section (\ref{sec:susy}), we show that these nine degenerate states form irreducible representations of the supersymmetric $spl(2,1)\otimes spl(2,1)$ algebra and provide the Clebsch-Gordon coefficients.

\subsection{US-US phase}

Now let us consider the case where the coupling strengths of both impurities are further decreased such that $a_{A,B}>3/2$. The impurities undergo phase transition from YSR phase into their respective unscreened (US) phases. At the phase transition point $a_{A,B}=3/2$, the energy of the bound state screening the respective impurity is exactly equal to the mass gap $\Delta$. As both the impurities enter the unscreened phase, these bound states `leak' into the bulk, thereby leaving the impurities unscreened. Hence, in the US-US phase, unlike the YSR-YSR phase, the impurities remain unscreened at all energies. The ground state in this phase $\ket{\rm{US';\!US'}}_{S^z_A,S^z_B}$ has even fermionic parity $\mathcal{P}=1$ and can be represented by 
\be \ket{\rm{US';\!US'}}_{S^z_A,S^z_B}\equiv \ket{S^z_A} \otimes \ket{S^z_B}.
\ee

The first and the second kets represent the z-components of the spin associated with the left and right boundaries respectively. Using this notation, we have
\be \ket{\rm{US';\!US'}}_{\pm 1/2,\pm 1/2}\equiv \ket{\pm \frac{1}{2}} \otimes \ket{\pm \frac{1}{2}}.
\ee
The two unscreened impurities give rise to a spin-$\frac{1}{2}$ at each boundary and hence they form an irreducible representation of the $SU(2) \otimes SU(2)$ algebra, which are the usual spin singlet and spin triplets of the two unscreened impurities.

The Kondo, YSR and unscreened phases exhaust the possibilities for a single impurity. Each impurity can be independently chosen to be in any of these, not just the same one as detailed already. For completeness we know describe the possible mixed phases.

\subsection{Kondo-YSR (K-YSR) phase}
The first of the mixed phases we consider is the case where the RG invariant parameter corresponding to the impurity $A$ takes real values in the range $d_A\in (0,\infty)$ or imaginary values in the range $a_A\in (0, 1/2)$, and that of the impurity $B$ takes imaginary values in the range $a_B\in (1/2,3/2)$. The impurity $A$, is screened by the Kondo effect with Kondo temperature $T^K_A$. The impurity $B$ is in its respective YSR phase. The state $\ket{\rm{K;\!S}}$ in which the impurity $B$ is screened by the localized bound state is a total singlet $\vec{S}_T=0$ and has even fermionic parity is $\mathcal{P}=+1$. Unlike the K-K phase, the single particle bound state localized near the impurity $B$ can be removed, and hence, one can unscreen the impurity $B$. The resultant state $\ket{\rm{K;\!US}}_{\pm /2}$, where the impurity $B$ is unscreened has a total spin $S^z=\pm 1/2$ and has odd fermionic parity $\mathcal{P}=-1$. Using the notation introduced above, this state is represented as
\begin{align}
\ket{\rm{K;\!S}}&\equiv \ket{0}\otimes \ket{1,0,0},\\
\ket{\rm{K;\!US}}_{\pm 1/2}& \equiv \ket{0}\otimes \ket{\frac{1}{2},\frac{1}{2}, \pm \frac{1}{2}}.
\end{align}
As usual, the first ket corresponds to the impurity $A$ which is screened by the Kondo effect and the second ket corresponds to impurity $B$. The energy of the bound state localized at the right boundary is again given by (\ref{Echarge}) with $A\rightarrow B$.

%equal to the difference of the energy $E_{\ket{\rm{K;\!S}},N}$ of the state $\ket{\rm{K;\!S}}$, which has even number of electrons $N$, and the energy $E_{\ket{\rm{K;\!US}},N\pm1}$ of the state $\ket{\rm{K;\!US}}$, which has odd number of electrons $N\pm1$ respectively\beE_{\text{Bound,B}}= E_{\ket{\rm{K;\!S}},N}- \frac{1}{2}(E_{\ket{\rm{K;\!US}}, N-1}+E_{\ket{\rm{K;\!US}},N+1}).\eeThe energy $E_B$ of this bound state has the charge and the spin part $E_{\text{Bound,B}}=E_{\text{charge}}+E_{\text{spin,B}}$, where \be\label{Echarge}E_{\rm charge} = -\frac{\pi}{2L}, \;\; E_{\text{spin,B}} = -\Delta \sin{\pi a_B}.\ee

 The ground state of the system is $\ket{\rm{K;\!S}}$ for $1/2<a_B<1$, whereas it is $\ket{\rm{K;\!US}}_{\pm 1/2}$ for $1<a_B<3/2$. For $1/2<a_B<1$, one can remove the bound state localized near the impurity $B$ from the ground state $\ket{\rm{K;\!S}}$ and obtain the state $\ket{\rm{K;\!US}}_{\pm 1/2}$, thereby unscreen the impurity $B$. For $1<a_B<3/2$, one can add this bound state to the ground state $\ket{\rm{K;\!US}}_{\pm 1/2}$ and obtain the state $\ket{\rm{K;\!S}}$ there by screen the impurity $B$. Just as before, both of these processes cost energy $E_{{\rm{bound}},B}$, which is always smaller than the mass gap $\Delta$. Hence, the state above the ground state is called the mid-gap state, which is $\ket{\rm{K;\!US}}_{\pm 1/2}$ for $1/2<a_B<1$ and $\ket{\rm{K;\!S}}$ for $1<a_B<3/2$. At $a_B=1$, $E_{\text{spin,B}}=0$, where the states $ \ket{\rm{K;\!S}}$ and  $\ket{\rm{K;\!US}}_{\pm 1/2}$ are degenerate, and a first order quantum phase transition occurs. Notice that in the thermodynamic limit $L\rightarrow \infty$, $E_{\text{charge}}\rightarrow 0$, and hence there exists a genuine zero-energy mode (ZEM) at $a_B=1$. 

Analogously, we can also consider the case where %where the coupling strength of the impurity $B$ takes real values in the range $d_B\in (0,\infty)$ or imaginary values in the range $a_B\in [0,1/2)$, whereas that of the impurity $A$ takes imaginary values in the range $a_{A}\in (1/2,3/2)$, 
impurity $B$ is in its Kondo phase whereas the impurity $A$ is in its YSR phase. Similar to the previous case, the state $\ket{\rm{S';\!K}}$ in which the impurity $A$ is screened by the bound state localized near the left boundary and the state $\ket{\rm{US';\!K}}_{\pm 1/2}$ in which the bound state is absent leaving the impurity $A$ is given by
 \begin{align}
\ket{\rm{S';\!K}}&\equiv  \ket{1,0,0} \otimes \ket{0},\\
\ket{\rm{US';\!K}}_{\pm 1/2}& \equiv  \ket{\frac{1}{2},\frac{1}{2}, \pm \frac{1}{2}} \otimes \ket{0}.
     \end{align}
The energy of the bound state localized near the impurity $E_{\text{Bound,A}}$ takes the same form as (\ref{Echarge}).

 \subsection{Kondo-US (K-US) phase}
Next, we examine the case where the coupling strength of impurity $A$ is such that it is in its respective Kondo phase, and the impurity coupling strength corresponding to impurity $B$ is decreased further such that $a_B>3/2$, where the impurity $B$ is in its unscreened (US) phase. At the phase transition point $a_B=3/2$, the energy of the state $\ket{\rm{K;\!S}}$ with respect to the ground state $\ket{\rm{K;\!US}}_{\pm 1/2}$ is exactly equal to the mass gap $\Delta$. As impurity $B$ enters its unscreened phase, as discussed above, the bound state leaks into the bulk, thereby leaving the impurity $B$ unscreened. Hence, in the K-US phase, unlike in the K-YSR phase, the impurity $B$ remains unscreened at all energies. The ground state in this phase $\ket{\rm{K;\!US}'}_{S^z_B}$ can be represented by 
\be \ket{\rm{K;\!US}'}_{S^z_B}\equiv \ket{0} \otimes \ket{S^z_B},
\ee
where as usual, the first ket represents impurity $A$ being screened by the Kondo effect and the second ket represents the z-component of the spin associated with the right boundary. Using this notation, we have
\be \ket{\rm{K;\!US}'}_{\pm 1/2}\equiv \ket{0} \otimes \ket{\pm \frac{1}{2}}.
\ee
This state has odd fermionic parity $\mathcal{P}=-1$. Similarly, for the case where the impurity $B$ is in its Kondo phase whereas the impurity $A$ is in its unscreened phase, the ground state of the system $\ket{\rm{US';\!K}}_{\pm 1/2}$ is represented by 
\be \ket{\rm{US';\!K}}_{\pm 1/2}\equiv  \ket{\pm \frac{1}{2}}\otimes \ket{0},
\ee
where the impurity $A$ is always unscreened at all energies. Just as in the previous case, this state has odd fermionic parity $\mathcal{P}=-1$.

\subsection{YSR-US}
Lastly, let us take the case where the impurity $A$ is in its YSR phase, whereas the impurity $B$ is in its unscreened phase. The state $\ket{\rm{S;\!US}'}_{\pm 1/2}$ in which the impurity $A$ is screened by the bound state and the state $\ket{\rm{US;\!US}'}_{\pm 1/2,\pm 1/2}$ in which the bound state screening the impurity $A$ is absent such that it is unscreened, while the impurity $B$ remains unscreened in both the states can be represented by

\begin{align}
\ket{\rm{S;\! US'}}_{\pm 1/2}&\equiv \ket{1,0,0} \otimes \ket{\pm \frac{1}{2}},\\
\ket{\rm{US;\!US}'}_{\pm 1/2,\pm 1/2}&\equiv \ket{\frac{1}{2},\frac{1}{2},\pm \frac{1}{2}}\otimes \ket{\pm \frac{1}{2}}.
\end{align}

These states $\ket{S, US'}_{\pm 1/2}, \ket{\rm{US;\!US}'}_{\pm 1/2,\pm 1/2}$ have odd $\mathcal{P}=-1$ and even $\mathcal{P}=1$ fermionic parities respectively. Similarly, in the case where 
impurity $B$ is in its YSR phase, whereas the impurity $A$ is in its unscreened phase. The state $\ket{US',S}_{\pm 1/2}$ in which the impurity $A$ is screened by the bound state and the state $\ket{US',US}_{\pm 1/2,\pm 1/2}$ in which the bound state screening the impurity $B$ is absent such that it is unscreened, while the impurity $A$ remains unscreened in both the states can be represented by

\begin{align}
\ket{{\rm{US', S}}}&\equiv  \ket{\pm \frac{1}{2}}\otimes \ket{1,0,0},\\
\ket{{\rm{US',US}}}&\equiv \ket{\pm \frac{1}{2}}\otimes \ket{\frac{1}{2},\frac{1}{2},\pm \frac{1}{2}}.
\end{align}

Similar to the previous case, these states $\ket{{\rm{US',S}}}_{\pm 1/2}, \ket{{\rm{US',US}}}_{\pm 1/2,\pm 1/2}$ have odd $\mathcal{P}=-1$ and even $\mathcal{P}=1$ fermionic parities respectively.

This completes the characterization of the low energy, boundary states for all parameter regimes of our model.

\section{Supersymmetric algebra}
\label{sec:susy}

In the previous section we have exhaustively characterized the low energy, boundary spectrum of the Hamiltonian,~\eqref{Hamiltonian}. Notably, we have found that when one or both of the impurities are in the YSR phase, there are either three or nine states in the low energy Hilbert space, which can all become degenerate at a particular set of values in the parameter space. In Table (\ref{table1}), these states are represented as the direct product of the three states corresponding to each boundary, which form a representation of the supersymmetric $spl(2,1) \otimes spl(2,1)$ algebra. Below we provide a brief description of this algebra and provide further details in the appendices.

For the supersymmetric algebra $spl(2,1)$, there exists two quantum numbers which can be used to specify a particular representation. These are, the baryon number, $b$ and the total spin quantum number, $q$.  Each representation, $[b,q]$, specified by a specific choice of these quantum numbers may contain up to four multiplets  $|b,q,q_3\rangle, \;\; |b+\frac{1}{2},q-\frac{1}{2},q_3\rangle, \;\; |b-\frac{1}{2},q-\frac{1}{2},q_3\rangle, \;\; |b,q-1,q_3\rangle$, where $q_3=-q,...,q$, is the spin $S^z$. Different representations corresponding to different values of $b,q$ may or may not have all of these multiplets \cite{Rittenberg}. 

\begin{figure}[!h]
\includegraphics[width=0.7\columnwidth]{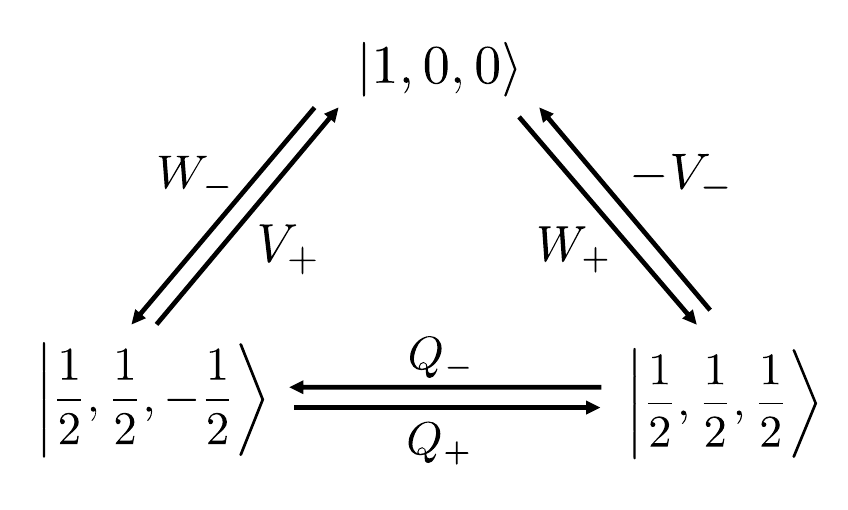}
\caption{The schematic shows the action of the generators of the supersymmetric $spl(2,1)$ algebra on the three states corresponding to the representation $[\frac{1}{2},\frac{1}{2}]$, which is associated with each boundary.}
\label{fig:irrep3}
\end{figure}

Amongst the generators of the algebra, there exist four operators, $V_\pm,W_\pm$, which connect different multiplets in a given representation, two operators $Q_\pm$ which connect different elements within a multiplet and two further operators, $B,Q_3$, which are diagonal. The operators $V_{\pm}$ increase the baryon number by $\frac{1}{2}$ and simultaneously increase/decrease the spin $S^z$ by $\frac{1}{2}$. The operators $W_{\pm}$, instead, decrease the baryon number by $\frac{1}{2}$ while simultaneously increasing/decreasing the spin $S^z$ by $\frac{1}{2}$. The operators $Q_{\pm}$ do not change the baryon number but only increase/decrease the spin $S^z$ by $1$. The remaining operators, $Q_3$ and $B$, measure the values of $q_3$ and $b$, respectively. The algebra has two Casimir elements, denoted $K_2,K_3$ whose explicit form is given in the appendix. They have eigenvalues $q^2-b^2$ and $b(q^2-b^2)$ when acting on the representation $[b,q]$. 

Comparing this to the notation used in (\ref{repphy}), we can identify  the quantity $\frac{n_{i,A}+n_{b,A}}{2}$, with the baryon number at the left edge, the total spin quantum number  $q$, with $S_{T,A}$ and $q_3$ 
 with $S^z_{A}$. Likewise, we can make the same identifications at the right edge with the replacement $A\to B$.

\begin{figure}[!h]
\includegraphics[width=1\columnwidth]{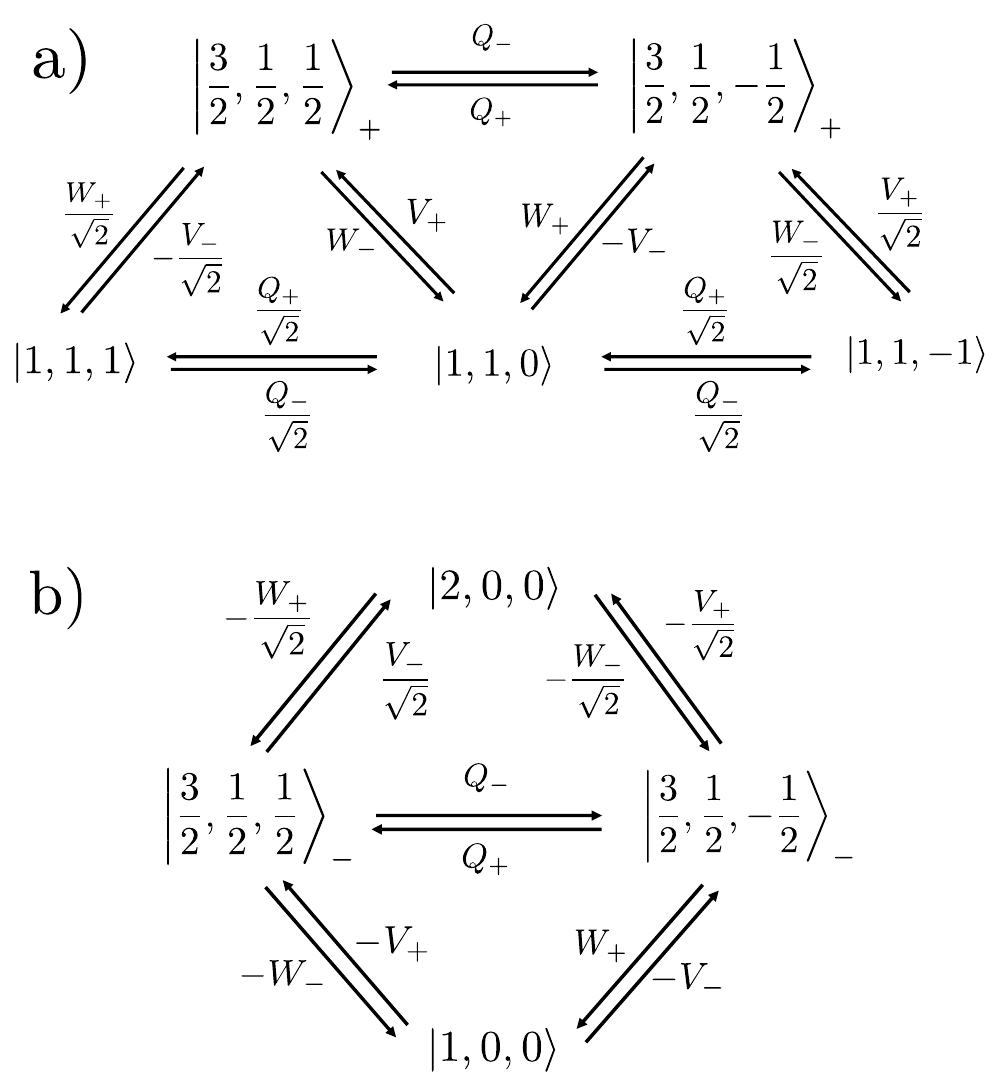}
\caption{The schematics a) and b) show the action of the generators of the supersymmetric $spl(2,1)$ algebra on the states corresponding to the irreducible representation $[1,1]$ and $[\frac{3}{2},\frac{1}{2}]$ respectively. The $[1,1]$ representation contains five states which are even under space inversion $L\leftrightarrow R$. The representation $[\frac{3}{2},\frac{1}{2}]$ contains four states, where except the state $\ket{2,0,0}$ which is even, the rest of the states are odd under space inversion $L\leftrightarrow R$.}
\label{fig:irrep}
\end{figure}

%\begin{figure}[!h]\includegraphics[width=0.7\columnwidth]{irrep2.pdf}\caption{P}\label{fig:irrep2}\end{figure}

Having identified the supersymmetric quantum numbers with those of our system we now identify which representation is realized at each edge. To do this we simply note that, at each edge, the total spin can be either $S^z=\pm \frac{1}{2}$ or $S=0$, meaning that the representation can only be $[\frac{1}{2},\frac{1}{2}]$ which contains the multiplets $|\frac{1}{2},\frac{1}{2},\pm \frac{1}{2}\rangle$ and $\ket{1,0,0}$. The former multiplet, $|\frac{1}{2},\frac{1}{2},\pm \frac{1}{2}\rangle$, correspond to the states in which the impurity is unscreened, whereas the latter, $\ket{1,0,0}$, corresponds to the state in which the impurity is screened by the localized bound state. These multiplets and their constituent elements are related to each other through the operators $Q_\pm,W_\pm,V_\pm$ as shown in Fig: (\ref{fig:irrep3}).

\begin{table}[h!]
\centering
\caption{The table provides explicit expressions for the irreducible representations $[1,1]$ and $[\frac{3}{2},\frac{1}{2}]$ (left column) of the nine degenerate ground states at the  supersymmetric point in terms of the direct product of the representations $[\frac{1}{2},\frac{1}{2}]$ associated with each edge (right column) along with the Clebsch-Gordon coefficients.}
\begin{tabular}{|c|c|}
\hline
\hline
 $[1,1]$& $[1/2,1/2]\otimes [1/2,1/2]$\\
 \hline $|\frac{3}{2},\frac{1}{2},\pm\frac{1}{2}\rangle_+$ & $\frac{1}{\sqrt{2}}\left(|1,0,0\rangle \otimes |\frac{1}{2},\frac{1}{2},\pm\frac{1}{2}\rangle + |\frac{1}{2},\frac{1}{2},\pm\frac{1}{2}\rangle \otimes |1,0,0\rangle\right)$\\
$|1,1,1\rangle$ & $|\frac{1}{2},\frac{1}{2},\frac{1}{2}\rangle \otimes |\frac{1}{2},\frac{1}{2},\frac{1}{2}\rangle$\\
$|1,1,0\rangle$ & $\frac{1}{\sqrt{2}}\left(|\frac{1}{2},\frac{1}{2},\frac{1}{2}\rangle \otimes |\frac{1}{2},\frac{1}{2},-\frac{1}{2}\rangle+|\frac{1}{2},\frac{1}{2},-\frac{1}{2}\rangle \otimes |\frac{1}{2},\frac{1}{2},\frac{1}{2}\rangle\right) $\\
$|1,1,-1\rangle$ & $|\frac{1}{2},\frac{1}{2},-\frac{1}{2}\rangle \otimes |\frac{1}{2},\frac{1}{2},-\frac{1}{2}\rangle$\\
\hline
\hline
$[3/2,1/2]$& $[1/2,1/2]\otimes [1/2,1/2]$\\
\hline

$|\frac{3}{2},\frac{1}{2},\pm\frac{1}{2}\rangle_{-}$ & $\frac{1}{\sqrt{2}}\left(|1,0,0\rangle \otimes |\frac{1}{2},\frac{1}{2},\pm\frac{1}{2}\rangle - |\frac{1}{2},\frac{1}{2},\pm\frac{1}{2}\rangle \otimes |1,0,0\rangle\right)$\\
$|1,0,0\rangle$ &  $\frac{1}{\sqrt{2}}\left(|\frac{1}{2},\frac{1}{2},\frac{1}{2}\rangle \otimes |\frac{1}{2},\frac{1}{2},-\frac{1}{2}\rangle-|\frac{1}{2},\frac{1}{2},-\frac{1}{2}\rangle \otimes |\frac{1}{2},\frac{1}{2},\frac{1}{2}\rangle \right)$ \\
$|2,0,0\rangle$ & $|1,0,0\rangle\otimes |1,0,0\rangle$\\

\hline
\hline
\end{tabular}
\label{table3}
\end{table}

We now consider both boundaries together, which, when they are in their respective YSR phases, provide nine states in the low energy Hilbert space. This leads to the representation
$[\frac{1}{2},\frac{1}{2}]\otimes [\frac{1}{2},\frac{1}{2}]$ which can be decomposed in terms of irreducible representations.  We find that 
\be\label{tensorp} [\frac{1}{2},\frac{1}{2}]\otimes [\frac{1}{2},\frac{1}{2}]= [1,1] \oplus [\frac{3}{2},\frac{1}{2}]. \ee  
On the right hand side, the total baryon number, $b$ corresponds to half of the total number of impurities and the bound states at both the edges
\be b=\sum_{\alpha=A,B}\frac{n_{i,\alpha}+n_{b,\alpha}}{2},\ee
 and therefore, can take values $1,3/2,2$ corresponding to state with no bound states, the state with one bound state, either at the left or the right edge, and the state with two bound states, one at each edge respectively. Likewise, the total spin $q$ refers to the total spin of both boundaries.

 Both of the, irreps in~\eqref{tensorp} contain several mutliplets. In particular,  $[1,1]$ contains the multiplets $|1,1,m\rangle, |\frac{3}{2},\frac{1}{2},m\rangle$, while $[\frac{3}{2},\frac{1}{2}]$ contains the multiplets $|\frac{3}{2},\frac{1}{2},m\rangle, |1,0,0\rangle$, $\ket{2,0,0}$. These various multiplets are related to the direct product of the multiplets corresponding to the representation $[\frac{1}{2},\frac{1}{2}]$ at each boundary through the relations displayed in Table (\ref{table3}).  We now describe the physical properties of these various multiplets. 

In the representation $[1,1]$, the multiplet $\ket{1,1,m}$ has total spin $S_T=1$ and is even under spatial inversion, i.e. exchanging $A\leftrightarrow B$. They correspond to the triplet states of the two unscreened impurities. The states,  $\ket{\frac{3}{2},\frac{1}{2},\pm \frac{1}{2}}_{+}$ are also even under spatial inversion and corresponds to the states in which one impurity is screened by the localized bound state, whereas the other impurity is unscreened. The subscript $+$ denotes that it is an even under inversion.   

In the representation $[\frac{3}{2},\frac{1}{2}]$, the state $\ket{2,0,0}$ is even under spatial inversion and corresponds to the state in which both the impurities are screened by the respective localized bound states. The states $\ket{\frac{3}{2},\frac{1}{2},\pm \frac{1}{2}}_{-}$, correspond to the states in which one impurity is screened by the localized bound state, whereas the other impurity is unscreened. The subscript, $-$ distinguishes them from the similar multiplet in the other irrep and denotes that this is odd under spatial inversion. Lastly, the state $\ket{1,0,0}$ is also even under inversion, and corresponds to the state in which both the impurities are unscreened and form a singlet. The multiplets within each representation $[1,1]$ and $[\frac{3}{2},\frac{1}{2}]$ are related to each other through the relations shown in Fig. (\ref{fig:irrep}).

\section{Low Energy Effective Hamiltonian and Exact ZEM}
In the YSR regime we can use the generators which we identified in the previous section to write the effective low energy Hamiltonian. For a single impurity, say $A$, this is given by
\begin{align} \nonumber H^{A}_{\text{eff}}=&\,\frac{1}{2}(E_{\rm gs}-E_{{\rm{bound}},A})\Big(Q_{-}^AQ_{+}^A+Q_{+}^AQ_{-}^A\Big)\\\nonumber &+(E_{\rm gs}+E_{{\rm{bound}},A})\Big(V_{+}^AW_{-}^A+W_{-}^AV_{+}^A\\\label{lowenergyHA}
&\hspace{2.5cm}-V_{-}^AW_{+}^A-W_{+}^AV_{-}^A\Big)
\end{align}
Here $E_{\rm gs}$ is the ground state energy and the operators appearing here are those of the $[\frac{1}{2},\frac{1}{2}]$ representation acting on the edge $A$. The operators $Q^A_{\pm}$ that change the total $z$-component of the spin while not changing the baryon number are exact zero-energy operators, as they act between the states that are exactly degenerate. Since the baryon number is associated with the number of bound states, in order to change the total baryon number, one needs to add or remove a boundary bound state. From (\ref{Echarge}), we see that the energy of the boundary bound state $E_{{\rm bound},A}$
has charging energy and spin energy. The charging energy is inversely proportional to the system size $\sim 1/L$,  whereas the spin energy depends on the respective RG invariant $a_{A}$ and is zero when $a_{A}=1$. When $a_A\neq 1$, this Hamiltonian has a broken supersymmetry, which is restored at $a_A=1$, strictly in the limit $L\rightarrow \infty$ where the charging energy is zero.  

Now, considering the case of two impurities, the effective low energy Hamiltonian can be written as
\begin{align}\nonumber&
H_{\text{eff}}=H^A_{\text{eff}}+H^B_{\text{eff}}.\end{align}
where $H^B_{\text{eff}}$ is the same as~\eqref{lowenergyHA} with $A\to B$ and where the operators with superscript $B$ act on that edge. We see that for $a_A,a_B\neq 1$, the Hamiltonian has broken supersymmetry. At $a_A=a_B=1$, the supersymmetry is restored strictly in the limit $L\rightarrow\infty$, as the charging energy associated with the bound states at either boundary vanishes, and hence we call this the \textit{supersymmetric point}. 

\subsection{Exact zero energy modes}
Having written the low energy Hamiltonian, we now construct the operators which create its zero energy modes at the supersymmetric point in the odd fermionic parity sector. Here, it is important to note the difference between the two impurity system and a semi-infinite one with only a single impurity.

In the case of one-impurity, there exists only one state with odd fermionic parity, which corresponds to the state in which the impurity is screened by the bound state. Removing the bound state changes the fermionic parity and costs at least the charging energy. Hence, strictly speaking, apart from the spin flip operators which flip the spin of the unscreened impurity in the even parity sector, there exist no zero energy modes in the one-impurity case. In contrast, for the case of two impurities, there exist four states in the odd fermionic parity sector, which correspond to one impurity being screened by a bound state whereas, the other impurity remains unscreened. Two of these states $|\frac{3}{2},\frac{1}{2},\pm\frac{1}{2}\rangle_+$
correspond to the representation $[1,1]$ and two other states $|\frac{3}{2},\frac{1}{2},\pm\frac{1}{2}\rangle_-$
correspond to the representation $[3/2,1/2]$ as shown in Table (\ref{table3}). There exist three ZEM which map the pair of states in one representation with the corresponding one in the other representation. These operators are given by

\begin{align}
\Sigma^x_{+}&= W_{+}^A V_{-}^B-W_{+}
^BV_{-}^A,\\\Sigma^x_{-}&=-i(W_{-}^AV_{+}^B-W_{-}
^BV_{+}^A),\\
\Sigma^z&=4B^A-3.
\end{align}
These operators act on the above mentioned states as follows
\begin{align}
\Sigma^x_{+}|\frac{3}{2},\frac{1}{2},\frac{1}{2}\rangle_{\pm}&=\pm |\frac{3}{2},\frac{1}{2},\frac{1}{2}\rangle_{\mp},  \\
\Sigma^x_{-}|\frac{3}{2},\frac{1}{2},-\frac{1}{2}\rangle_{\pm}&=\pm i |\frac{3}{2},\pm\frac{1}{2},-\frac{1}{2}\rangle_{\mp}, \\\Sigma^z |\frac{3}{2},\frac{1}{2},\pm\frac{1}{2}\rangle_{\pm}&= |\frac{3}{2},\frac{1}{2},\pm\frac{1}{2}\rangle_{\mp}.\end{align}
At the supersymmetric point, apart from the trivial spin flip operators, the above mentioned operators act on the states that are odd under fermionic parity and are exponentially degenerated, i.e.  to the  exponential accuracy in the system size $e^{-L}$. Hence, as opposed to the case of one impurity, they do not require charging energy. Thus, we see that when both boundaries of a spin singlet superconductor are coupled to magnetic impurities, as opposed to just one boundary, the system exhibits exact ZEM.  Up until now, we have discussed exact ZEM in the odd fermionic parity sector. Similarly, one can construct ZEM that act in the even fermionic parity sector. Unlike the ZEM in odd fermionic parity sector discussed above, the ZEM in the even fermionic parity sector involve the  charging energy which scales as $1/L$, and hence, they are not exact ZEM.

\section{Discussion}
\label{sec:conclusion}
We have considered a one-dimensional spin-singlet superconductor and coupled the two edges to spin 1/2 magnetic impurities. The electrons at the edges interact with the impurities through a spin- exchange interaction. In the absence of the impurities, the spin-singlet superconductor exhibits a trivial phase with a unique ground state. When coupled to the impurities, the system exhibits nine different phases depending on the coupling strengths associated with both impurities, which are labeled by the phase exhibited by the impurity at the left boundary followed by the phases exhibited by the impurity at the right boundary. There exist three phases associated with each impurity: Kondo phase, YSR phase and unscreened phase. In the Kondo phase, the impurity is screened by the Kondo effect, whereas in the YSR phase, the impurity is screened by a bound state localized near the impurity whose energy is less than the superconducting gap $\Delta$. This bound state can be removed, and thereby the impurity can be unscreened. In the unscreened phase, this bound state ceases to exist, and hence the impurity remains unscreened. Considering both boundaries, in the phase where both the impurities are in their respective YSR phases, there exist two bound states, one corresponding to each edge.  At the supersymmetric point, both these bound states carry zero energy and the system gives rise to a nine-fold degenerate ground state. These nine states form irreducible representations of the supersymmetric $spl(2,1)\otimes spl(2,1)$ algebra giving rise to exact ZEM. 

Our results raise several important questions. The first is whether the rich phase diagram exhibited by the system is associated with an SPT phase. The second is whether the nine degenerate ground states at the supersymmetric point exhibit local boundary quantum numbers similar to the fractional boundary quantum numbers \cite{Jackiw,JackiwRebbi,Rajaraman} exhibited by SPTs  and SSBs \cite{PAA1,XXZpaper}. To answer the first one, one needs to identify a non-local order parameter or look at whether the system exhibits degeneracy in the entanglement spectrum.  To answer the second question, one needs to analyze the spin-profiles in the nine degenerate ground states and look into the possible existence of local quantum numbers at the boundaries. To tackle these questions, one can employ tensor network methods such as DMRG
and study the effect of magnetic impurities coupled to the edges of a lattice model \cite{Keselman2015} associated with the spin-singlet superconductor.

\acknowledgements
CR acknowledges support from the  European Union - NextGenerationEU, in the framework of the PRIN Project HIGHEST code 2022SJCKAH. 

\bibliography{refks}

\begin{widetext}

\appendix
\section{Bethe Ansatz}
\subsection{Hamiltonian}

The Hamiltonian of the system is

\be
\label{HamiltonianGNK2}
H= H_{\text{GN} }+ H_{\text{imp}},
\ee
where $H_{\text{GN}}  = \int_{-L/2}^{L/2} dx\;  \mathcal{H}_{\text{GN}} $
is the Hamiltonian of the Gross-Neveu (GN) model with
\bea
\label{GN2}
 \mathcal{H}_{\text{GN}}&=&-i (\psi^{\dagger}_{R a}\partial_x \psi^{}_{R a}-\psi^{\dagger}_{L a}\partial_x \psi^{}_{L a})-  \\
 &-&2g \psi^{\dagger}_{Ra}\psi^{\dagger}_{Lc}\left(\sigma^x_{ab}\sigma^x_{cd}+\sigma^y_{ab}\sigma^y_{cd}+\sigma^z_{ab}\sigma^z_{cd}\right)\psi_{Rb}\psi_{Ld} \nonumber,
\eea
and 
\be
\label{kondo2}
H_{\text{imp}}= -J_A\vec{\sigma}_{ab}\cdot\vec{S}_{\alpha\beta}\psi^{\dagger}_{La}(-L/2)\psi^{}_{Rb}(-L/2)-J_B\vec{\sigma}_{ab}\cdot\vec{S}_{\alpha\beta}\psi^{\dagger}_{La}(L/2)\psi^{}_{Rb}(L/2),
\ee

where $J_A,J_B$ are the coupling strengths of the two impurities at the left and the right edges respectively.
\vspace{3mm}

\subsection{N-particle solution}

The Hamiltonian commutes with total particle number,\newline
$N=\int \psi_+^\dag(x)\psi_+(x)+\psi_-^\dag(x)\psi_-(x)$ and $H$ can be diagonalized by constructing the exact eigenstates in each $N$ sector. 

Since $N$ is a good quantum number we may construct the eigenstates by examining the different $N$ particle sectors separately. We start with $N=1$ wherein we can write the wavefunction as an expansion in plane waves,

\begin{eqnarray}\nonumber
\ket{k}=\sum_{a_j=\uparrow\downarrow,\sigma=\pm}\int_{-\frac{L}{2}}^{\frac{L}{2}}\mathrm{d}x\, e^{i\sigma kx} A^\sigma_{a_1a_Aa_B}    \psi^{\dagger}_{\sigma,a_1} (x)\ket{0} \otimes \ket{a_A}\otimes \ket{a_B}.
\end{eqnarray}

 $\ket{0}$ is the drained Fermi sea and $A^\sigma_{a_1a_Aa_B}$  are the amplitudes for an electron with chirality $\sigma$ and spin $a_1$ and the impurity $A$ and $B$ having spins $a_A$ and $a_B$ respectively. The two boundary S-matrices $S^{1R}_{a_1b_1,a_Bb_B},S^{1L}_{a_1b_1,a_Ab_A}$ exchange the chirality of a particle. 
\bea A^-_{a_1a_Aa_B}=S^{1B}_{a_1b_1,a_Bb_B} \; A^+_{b_1a_Ab_B} \\A^+_{a_1a_Aa_B}=S^{1A}_{a_1b_1,a_Ab_A} \; A^-_{b_1b_Aa_B}.\eea

where 
\begin{eqnarray}\label{appensimp}
S^{1m}_{a_1b_1,a_jb_j}=e^{i\gamma} \; \left(\frac{\; ic_m I^{1m}_{a_1b_1,a_mb_m}+P^{1m}_{a_1b_1,a_mb_m}}{ic_m+1}\right),\\\label{c} c_m=\frac{1-3J_m^2/4}{2J},~e^{i\gamma_m}=\frac{3J_m^2-1+2iJ_m}{3J_m^2-1+iJ_m}.\end{eqnarray}

 Here $P^{1m}$ is the usual permutation operator $P^{1m}_{a_1b_1,a_mb_m}=(I^{1m}_{a_1b_1,a_mb_m}+ \vec{\sigma}^1_{a_1b_1}\cdot\vec{\sigma}^m_{a_mb_m})/2$ which exchanges the spins of particle 1 and impurity m, {{and $I^{1m}$ is the identity}}. The superscripts refer to the particle space on which this operator acts i.e. $1$ refers to the particle and $m$ to the impurity, $m=A,B$. Applying the boundary conditions also quantizes the bare particle momentum $k$. We now consider the two particle sector, $N=2$, were the bulk interaction plays a role.

Since the two particle interaction is point-like
we may divide configuration space into regions such that
the interactions only occur at the boundary between two
regions. Therefore away from these boundaries we write
the wave function as a sum over plane waves so that the most general two particle state can be written as
\bea\nonumber
\ket{k_1,k_2}= \sum_{\sigma,a} \int_{-\frac{L}{2}}^{\frac{L}{2}}\mathrm{d}^2x\,F_{a_1a_2a_Aa_B}^{\sigma_1\sigma_2}(x_1,x_2)e^{\sum_{j=1}^2i\sigma_jk_jx_j}\\\psi^{\dagger}_{\sigma_1a_1}(x_1)\psi^{\dagger}_{\sigma_2a_2}(x_2) \ket{0} \otimes \ket{a_A}\otimes \ket{a_B},
\label{2particle}\eea 
where we sum over all possible spin and chirality configurations  and the two particle wavefunction, $F_{a_1a_2a_Aa_B}^{\sigma_1\sigma_2}(x_1,x_2)$ is split up according to the ordering of the particles,
\bea
 F_{a_1a_2a_Aa_B}^{\sigma_1\sigma_2}=A_{a_1a_2a_Aa_B}^{\sigma_1\sigma_2}[12]\theta(x_2-x_1)+A_{a_1a_2a_Aa_B}^{\sigma_1\sigma_2}[21]\theta(x_1-x_2).
\eea

The amplitudes $A_{a_1a_2a_Aa_B}^{\sigma_1\sigma_2}[Q]$ refer to a certain chirality and spin configuration, specified by $\sigma_j$, $a_j$ as well as an ordering of the particles in configuration space denoted by $Q$. For $Q=12$ particle $1$ is to the left of particle $2$ while for $Q=21$ the  order of the particles are exchanged. Applying the Hamiltonian to \eqref{2particle} we find that it is an eigenstate with energy $E=k_1+k_2$ provided that these amplitudes are related to each other via application of $S$-matrices. The amplitudes which differ by exchanging the chirality of the leftmost or the rightmost particle are related by the boundary S-matrices.

\bea
A^{\sigma_1-}[12]=S^{2B} \;A^{\sigma_1+}[12], \;\; A^{+\sigma_2}[12]=S^{1A} \;A^{-\sigma_2}[12],\\
A^{-\sigma_2}[21]=S^{1B}\;A^{+\sigma_2}[21],\;\; A^{\sigma_1+}[21]=S^{2A}\;A^{\sigma_1-}[21].\eea
 
 The one particle boundary S-matrices are given by \ref{appensimp}. For ease of notation we have suppressed spin indices. It is understood that $S^{1B},S^{1A}$ act in the spin space of particle 1 whereas $S^{2B},S^{2A}$ act in the spin space of particle 2.

There are two types of two particle bulk $S$-matrices denoted by $S^{12}$ and $W^{12}$ which arise due to the bulk interactions and relate amplitudes which have different orderings. The first relates amplitudes which differ by exchanging the order of particles with opposite chirality 
\bea
A^{+-}[21]=S^{12}A^{+-}[12],\\
A^{-+}[12]=S^{12}A^{-+}[21],
\eea
where  $S^{12}$ acts on the spin spaces of particles 1 and 2. Explicitly it is given by

\begin{eqnarray}\label{S12}S^{12}_{a_1a_2,b_1b_2}= e^{i\phi}\; \left(\frac{2ib \; I^{12}_{a_1a_2,b_1b_2} +P^{12}_{a_1a_2,b_1b_2}}{2ib+1}\right),~~~\\\label{b}
b= \frac{1-3g^2/4}{4g},~e^{i\phi}=\frac{3g^2-1+2ig}{3g^2-1+ig}. \end{eqnarray}

Whilst the second {{type of $S$-matrix}} relates amplitudes where  particles of the same chirality are exchanged,
\bea
A^{--}[21]=W^{12}A^{--}[12],\\\label{W12}
A^{++}[12]=W^{12}A^{++}[21].
\eea
 Unlike \eqref{S12}, $W^{12}$ is not fixed by the Hamiltonian but rather by the consistency of the construction. This is expressed through the Yang-Baxter equations 
\bea\label{BYB1}
S^{23}\;S^{13}\;W^{12}&=&W^{12}\;S^{13}\;S^{23},\\
W^{23} \;W^{13} \;W^{12}& = &W^{12} \;W^{13}\; W^{23},\\  S^{2B}\;S^{12}\;S^{1B}\;W^{12}&=&W^{12}\;S^{1B}\;S^{12}\;S^{2B},\\\label{BYB2}S^{2A}\;S^{12}\;S^{1A}\;W^{12}&=&W^{12}\;S^{1A}\;S^{12}\;S^{2A},\eea

which need to be satisfied for the eigenstate to be consistent. We take $W^{12}=P^{12}$ which can be explicitly checked to satisfy \eqref{BYB1}-\eqref{BYB2}. The relations \eqref{2particle}-\eqref{W12} provide a complete set of solutions of the two particle problem.

We can now generalise this to the $N$-particle sector and find that the eigenstates of energy $E=\sum_{j=1}^Nk_j$ are of the form
\begin{eqnarray}\label{NparticleS}
\ket{\{k_j\}}=
\sum_{Q,\vec{a},\vec{\sigma}}\int \theta(x_Q) A^{\{\sigma\}}_{\{a\}}[Q] \prod_j^{N_e} e^{i\sigma_j k_jx_j}\psi^{\dagger}_{a_j\sigma_j}(x_j)\ket{0}
 \otimes \ket{a_A}\otimes \ket{a_B}.
\end{eqnarray}

Here we sum over all  spin and chirality configurations specified by
\newline $\{a\}=\{a_1\dots a_Na_Aa_B\}$, $\{\sigma\}=\{\sigma_1\dots \sigma_N\}$ as well as different different orderings of the $N$ particles. These different orderings correspond to elements of the symmetric group $Q\in \mathcal{S}_N$. In addition $\theta(x_Q)$ is the Heaviside function which is nonzero only for that particular ordering. 
As in the $N=1,2$ sectors the amplitudes $A^{\vec{\sigma}}_{\vec{a}}[Q]$ are related to each other by the various $S$-matrices in the same manner as before i.e. amplitudes which differ by changing the chirality of the rightmost particle are related by the impurity $S$-matrix, $S^{j0}$, amplitudes which differ by exchanging the order of opposite or same chirality particles are related by $S^{ij}$ and $W^{ij}$ respectively. The consistency of this construction is then guaranteed by virtue of these $S$-matrices satisfying the following  reflection and  Yang-Baxter equations\cite{Sklyannin, Cherednik, ZinnJustin}
\bea\label{YB1}
W^{jk} \;W^{ik}\; W^{ij} &=& W^{ij} \;W^{ik} \;W^{jk},\\ \label{YB2}
S^{jk}\;S^{ik}\;W^{ij} &=& W^{ij}\;S^{ik}\;S^{jk},
\\ S^{jB}\;S^{ij}\;S^{iB}\;W^{ij}&=&W^{ij}\;S^{iB}\;S^{ij}\;S^{jB}\label{YB3},\\S^{jA}\;S^{ij}\;S^{iA}\;W^{ij}&=&W^{ij}\;S^{iA}\;S^{ij}\;S^{jA}\label{YB4},\eea
 Where $W^{ij}=P^{ij}$ and as before the superscripts denote which particles the operators act upon.

\subsection{Bethe equations}

Enforcing the boundary conditions on the eigenstate \eqref{NparticleS} we obtain the following eigenvalue problem which constrains the $k_j$,

\begin{eqnarray}
e^{-2ik_jL}A^{\{\sigma\}}_{\{a\}}[\mathbb{1}]=\left(Z_j\right)^{\{\sigma\},\{\sigma\}'}_{\{a\},\{a\}'} A[\mathbb{1}]^{\vec{\sigma}'}_{\vec{a}'}.
\end{eqnarray}
Here $\mathbb{1}$ denotes the identity element of $\mathcal{S}_N$, i.e. $\mathbb{1}=12\dots N$ and the operator $Z_j$ is the transfer matrix for the $j^\text{th}$ particle given by
\begin{eqnarray}
Z^j=W^{jj-1}\dots W^{j1} S^{jA}S^{j1}...S^{jj-1}S^{jj+1}...S^{jN}S^{jB}W^{jN}...W^{jj+1}\end{eqnarray}
where the spin indices have been suppressed. This operator takes the $j^\text{th}$ particle from one side of the system to the other and back again, picking up $S$-matrix factors along the way as it moves past the other $N-1$ particles, first as a right mover and then as a left mover.   Using the relations \eqref{YB1} \eqref{YB2} and \eqref{YB3}, one can prove that all the transfer matrices commute, $[Z_j,Z_k]=0$ and therefore are simultaneously diagonalizable. In order to determine the spectrum of $H$ we must therefore diagonalize $Z_j,~\forall j$. Here we choose to diagonalize $Z_1$. To do this we use the method of boundary algebraic Bethe Ansatz \cite{Sklyannin, Cherednik, ODBA}.  In order to use this method we need to embed the bare S-matrices in a continuum that is, we need to find the matrices $R(\lambda)$, $K(\lambda)$ such that for certain values of the spectral parameter $\lambda$, we obtain the bare S-matrices of our model.  $R(\lambda)$ turns out to be the $R$-matrix of the XXX spin chain which is given by \small{\begin{equation}R^{ij}_{ab}(\lambda)=\frac{1}{i\lambda+1}\left(i\lambda I^{ij}_{ab} + P^{ij}_{ab}\right).\end{equation}} We can see that \small{$R^{ij}(0)=W^{ij}, \hspace{2mm} R^{ij}(2b)= S_{ij}$}. We have ignored the unimportant constant $e^{i\phi}$. The transfer matrix{{ $Z_1$ }}is related to the Monodromy matrix $\Xi_{\tau}(\lambda)$ as $Z_1=t(b)=\Tr_{\tau} \Xi_{\tau}(b)$, where
\begin{equation}
\Xi_{\tau}(\lambda)=T_{\tau}(\lambda)\hat{T}_{\tau}(\lambda)
\end{equation}

with
\bea 
T_{\tau}(\lambda)&= &R_{1\tau}(\lambda+b)R_{A\tau}(\lambda+d_A)R_{2\tau}(\lambda+b)...R_{N\tau}(\lambda+b)R_{B\tau}(\lambda+d_B)\\
\hat{T}_{\tau}(\lambda)&=&R_{B\tau}(\lambda-d_B)R_{N\tau}(\lambda-b)...R_{2\tau}(\lambda-b)R_{A\tau}(\lambda-d_A)R_{1\tau}(\lambda-b).
\eea

Here $d_m= \sqrt{b^2-2b/c_m-1}$, $S^{1m}=R_{1m}(b-d_m)R_{1m}(b+d_m)$ and $m=A,B$. $\tau$ represents auxiliary space and $\Tr_\tau$ represents the trace in the auxiliary space. Using the the properties of the $R$ matrices one can prove that $[t(\lambda),t(\mu)]=0$ \cite{ODBA} and by expanding $t(\mu)$ in powers of $\mu$, obtain infinite set of conserved charges which guarantees integrability. By following the Boundary Algebraic Bethe Ansatz approach \cite{ODBA}, we obtain the Bethe equations 

%\bea\label{Bae}
%\left(\frac{\lambda_\alpha- b+i/2}{\lambda_\alpha- b-i/2}\right)^{N}\left(\frac{\lambda_\alpha+b+i/2}{\lambda_\alpha+ b-i/2}\right)^{N}\Pi_{j}^{A,B}\left(\frac{\lambda_\alpha-d_j+i/2}{\lambda_\alpha-d_j-i/2}\right)\left(\frac{\lambda_\alpha+d_j+i/2}{\lambda_\alpha+d_j-i/2}\right)=\Pi_{\alpha\neq \beta}^{M}\left(\frac{\lambda_\alpha-\lambda_\beta+i}{\lambda_\alpha-\lambda_\beta-i}\right)\left(\frac{\lambda_\alpha+\lambda_\beta+i}{\lambda_\alpha+\lambda_\beta-i}\right)\eea

\vspace{2mm}

%\bea\label{Ebae}
%e^{-2ik_jL} = \Pi_{\alpha=1}^M\left(\frac{\tilde b+\lambda_\alpha+i/2}{\tilde b+\lambda_\alpha-i/2}\right)\left(\frac{\tilde b-\lambda_\alpha+i/2}{\tilde b-\lambda_\alpha-i/2}\right)\eea

 \begin{eqnarray}\label{energysup}
e^{-ik_jL}\!=\!\prod_{\alpha=1}^Mf(2b, 2\lambda_\alpha),~
f(x,z)=\!\prod_{\sigma=\pm}\frac{x+\sigma z+i}{x+\sigma z-i}
\end{eqnarray}
where $\lambda_\alpha$, $\alpha=1,\dots,M$ satisfy the Bethe Ansatz equations
\begin{eqnarray}\label{BAEsup}
\prod_{m=A,B}\left[f(2\lambda_\alpha,2b)\right]^{N_e}f(2\lambda_{\alpha},2d_m)=\prod_{\alpha\neq \beta }^Mf(\lambda_\alpha,\lambda_\beta),
\end{eqnarray}

 These Bethe equations correspond to Bethe reference state with all up and all down spins. The Bethe roots govern the spin degrees of freedom of the system and $M\leq N/2$ gives the total $z$-component of spin, $S^z=N/2-M$. The solutions $\lambda_\alpha$ can be real or take complex values in the form of strings. In order to have a non vanishing wavefunction they must all be distinct, $\lambda_\alpha \neq \lambda_\beta$. In addition, the value $\lambda_\alpha=0$ should also be discarded as it results in a vanishing wavefunction \cite{ODBA}. Bethe equations of the type \eqref{BAEsup} are reflective symmetric, that is they are invariant under $\lambda_\alpha\rightarrow -\lambda_\alpha$ transformation. Due to this symmetry, solutions to the Bethe equations occur in pairs $\{-\lambda_\alpha,\lambda_\alpha\}$. 

\smallskip

 Applying logarithm to \eqref{BAEsup} we obtain 

\begin{eqnarray}\nonumber
-\pi I_j+\sum_{\sigma=\pm} N_e\Theta(\lambda_\alpha+\sigma b,1/2)+\sum_{m=A,B}\Theta(\lambda_\alpha+\sigma d_m ,1/2)\\+\Theta(\lambda_\alpha,1/2)+ \Theta(\lambda_\alpha,1)\label{Logbaesup}
=\sum_{\beta=-M}^{M} \Theta\left(\lambda_\alpha- \lambda_\beta,1\right)\end{eqnarray}

Where $\Theta(x,n)=\text{arctan}[x/n]$. And likewise taking the logarithm of \eqref{energysup} we get
\begin{eqnarray}\label{logEnergysup}
k_j=\frac{2\pi n_j}{L}+\frac{2}{L}\sum_{\beta=-M}^M\Theta( b -\lambda_\beta,1/2)
\end{eqnarray}

The integers $n_j$ and $I_\alpha$ arise from the logarithmic branch and serve as the quantum numbers of the states. The quantum numbers $I_\alpha$ correspond to the spin degrees of freedom while the quantum numbers $n_j$ are associated with the charge degrees of freedom and they must all be different. $I_\alpha$ and $n_j$ can be chosen independently implying the charge spin decoupling. Minimizing the ground state energy results in a cutoff such that $\pi|n_j|/L < \pi D$ where $D=N/L$ is the density \cite{AndreiLowenstein79}.  We consider here $b>0$. The model exhibits several phases depending on the values of $b$, $c_A$ and $c_B$ or equivalently $d_A$ and $d_B$.  $d_m$, $m=A,B$ can be real or purely imaginary, in which case we take $d_m=ia_m$. Below we will solve the Bethe equations and construct the ground state separately in each phase.

\section{Solution to the Bethe equations}
In this section, we provide the solution to the Bethe equations obtained above. We shall do so separately in each phase which correspond to particular values of the RG invariant parameters $d_A$ and $d_B$.

\subsection{Kondo-Kondo phase}
The \textit{Kondo-Kondo} (K-K) phase occurs for values of $d_A,d_B$ both real or both imaginary with $a_A,a_B<\frac{1}{2}$. In this phase both impurities are in Kondo phase. Consider first the case where $d_A,d_B$ are real.  The ground state is given by the particular choice of charge and spin quantum numbers $n^0_j$ , $I^0_\alpha$, where $n^0_j$ are consecutively filled from the lower cutoff $-LD$ upwards, and the integers $I^0_\alpha$  take consecutive values which corresponds to real valued  $\lambda_\alpha$ roots. In the limit $N\rightarrow \infty$ the Bethe roots fill the real line and the ground state can be described by $\rho(\lambda)$ the density of solutions $\lambda$, from which the properties of the ground state can be obtained. Reflection symmetry of the Bethe equations \eqref{BAEsup} allows us to define $\lambda_{-\alpha}=-\lambda_\alpha, \; \lambda_0=0$ \cite{XXXkondo} and introduce the counting function $\nu(\lambda)$ such that $\nu(\lambda_\alpha)=I_\alpha$.

Differentiating \eqref{Logbaesup}, and using $\rho(\lambda)=\frac{d}{d\lambda}\nu(\lambda)$ \cite{trieste}, we obtain the following integral equation

\begin{eqnarray}\label{gsrdensity}
g_{sr}(\lambda)&=&\rho_\text{sr}(\lambda)+\int_{-\infty}^{\infty}\mathrm{d}\mu\,\varphi(\lambda-\mu,1)\rho_\text{s}(\mu),
\end{eqnarray}

where $g_{sr}(\lambda)=\sum_{\sigma=\pm}N_e\varphi(\lambda+\sigma b ,1/2) +\sum_{m=A,B}\varphi(\lambda+\sigma d_m ,1/2)+\varphi(\lambda,1/2)+\varphi(\lambda,1)$ and  $\varphi(x,n)= (n/\pi)(n^2+x^2)^{-1}.$

Solving \eqref{gsrdensity} by Fourier transform we obtain the following Fourier transformed density distribution of Bethe roots 

\bea \label{denscreened1}\tilde\rho_{sr}(\omega)= \frac{N_e \cos[ b\,\omega] + \sum_{m=A,B}\cos[d_m \,\omega]+ \frac{1}{2}e^{-\frac{|\omega|}{2}}+\frac{1}{2}}{\sqrt{2\pi}\cosh[\frac{\omega}{2}]}.\eea

 Each of the terms here may be identified with a certain component of the system. The term which is proportional to $N$ is the contribution of the left and right moving electrons, the next term which depends on $d$ is the contribution due to the impurity, while the remaining terms can be associated with the boundaries at $x=0,-L/2$.  The number of Bethe roots $M_{sr}$ in the ground state of the Kondo phase  for $d$ real is given by
\bea
2M_{sr}+1=\int_{-\infty}^{+\infty}d\lambda \; \rho_{sr}(\lambda).
\label{nrootssr}
\eea
Using this the $z$-component of spin $(S^z)_{sr}$ of the ground state in this region can be obtained: $S^z_{sr}=N/2-M_{sr}$, $N=N_e+2$.  Using  $\tilde\rho(0)=\int\mathrm{d}\lambda\, \rho(\lambda)$ along with \eqref{denscreened1} we find that  \bea 
 (S^z)_{sr}=0.
 \eea 
 
Defining the fermionic parity as $\mathcal{P}=(-1)^{N_e}$ we find $\mathcal{P}=1$.  For the case of imaginary $d_m$ with $a_m<1/2$, we have the following logarithmic form of the Bethe equations from \eqref{BAEsup}
 \begin{flalign}\nonumber
\sum_{\sigma=\pm,m=A,B} N_e\Theta(\lambda_\alpha+\sigma b,1/2)+\Theta(\lambda_\alpha,a_m+1/2)+\Theta(\lambda_\alpha,1/2-a_m)\\+\Theta(\lambda_\alpha,1) \label{Logbae2}
=\sum_{\beta=-M}^{M} \Theta\left(\lambda_\alpha- \lambda_\beta,1\right) +\pi I_j. \end{flalign}

Differentiating \eqref{Logbae2} and following the same procedure as before we obtain

\begin{eqnarray}\label{gsidensity}
g_{si}(\lambda)&=&\rho_\text{si}(\lambda)+\int_{-\infty}^{\infty}\mathrm{d}\mu\,\varphi(\lambda-\mu,1)\rho_\text{s}(\mu),
\end{eqnarray}

where $g_{si}(\lambda)=\sum_{\sigma=\pm}N_e\varphi(\lambda+\sigma b ,1/2) +\sum_{m=A,B}(\varphi(\lambda, a_m+1/2)+\varphi(\lambda, 1/2-a_m))+\varphi(\lambda,1/2)+\varphi(\lambda,1)$.

Solving \eqref{gsidensity} by Fourier transform we obtain the following Fourier transformed density of roots

\bea  \label{denscreened2}\tilde\rho_\text{si}(\omega)= \frac{N_e \cos[ b\,\omega] + \sum_{m=A,B}\cosh[a_m\;\omega]+ \frac{1}{2}e^{-\frac{|\omega|}{2}}+\frac{1}{2}}{\sqrt{2\pi}\cosh[\frac{\omega}{2}]}.\eea

As can be seen, the above result can be obtained by analytical continuation of the result for $d_A,d_B$ real. The number of roots is given by formula same as \eqref{nrootssr}, from which we obtain the $z$-component of the spin of the ground state of the K-K phase for imaginary $d_A,d_B$ as \bea(S^z)_{si}=0.\eea Fermionic parity of this state is again $\mathcal{P}=1$.  The ground state in this phase, which we represent by $\ket{\rm{S;\!S}}$, is unique and is described by the distribution $\tilde\rho_\text{sr}(\omega)$ for $d_A,d_B$ real and by the distribution $\tilde\rho_\text{si}(\omega)$ for imaginary $d_A,d_B$ with $a_A,a_B<1/2$. Hence the ground state $\ket{\rm{S;\!S}}$ in the K-K phase has total spin $\vec{S}_T=0$ with even fermionic parity. Both the impurities are completely screened by the electrons in the bulk as a result of the Kondo effect. This can be confirmed by calculating the ratio of the density of states of the impurities and that of the bulk which has a lorentzian peak characteristic of the Kondo effect.
 
 \smallskip
 
 \subsection{Kondo-YSR phase}
The \textit{YSR-Kondo} (YSR-K) phase occurs for $d_A$ real or imaginary with $a_A<\frac{1}{2}$ and $d_B$ imaginary with $\frac{1}{2}<a_B<\frac{3}{2}$. In this phase the impurity at the left boundary is in YSR phase, where as the impurity at the right boundary is in its Kondo phase. There are two sub-phases $\text{(YSR-K)}_1$ and $\text{(YSR-K)}_2$ which occur for $\frac{1}{2}<a_B<1$ and $1<a_B<\frac{3}{2}$ respectively.
 
\smallskip

 Consider the state with all real $\lambda_\alpha$. Applying logarithm to the Bethe equations \eqref{BAEsup} we obtain

\begin{flalign} \nonumber\sum_{\sigma=\pm} N_e\Theta(\lambda_\alpha+\sigma b,1/2)+\Theta(\lambda_\alpha,a_B+1/2)-\Theta(\lambda_\alpha,a_B-1/2)\\+\Theta(\lambda_\alpha,a_A+1/2)+\Theta(\lambda_\alpha,1/2-a_A)+\Theta(\lambda_\alpha,1) 
=\sum_{\beta=-M}^{M} \Theta\left(\lambda_\alpha- \lambda_\beta,1\right) +\pi I_j. \label{Logbae3}\end{flalign}
 
 Differentiating \eqref{Logbae3} and following the same procedure as before we obtain

\begin{eqnarray}\label{gmdensity}
g_{usB}(\lambda)&=&\rho_\text{us}(\lambda)+\int_{-\infty}^{\infty}\mathrm{d}\mu\,\varphi(\lambda-\mu,1)\rho_\text{s}(\mu).
\end{eqnarray}

where $g_{usB}(\lambda)=\sum_{\sigma=\pm}N_e\varphi(\lambda+\sigma b ,1/2) +\varphi(\lambda, a_B+1/2)-\varphi(\lambda, a_B-1/2)+\varphi(\lambda, a_A+1/2)+\varphi(\lambda, 1/2-a_A)+\varphi(\lambda,1/2)+\varphi(\lambda,1)$.

 Solving \eqref{gmdensity} by Fourier transform we obtain the following Fourier transformed density distribution of roots

\bea  \label{denscreened3}\tilde\rho_\text{usB}(\omega)= \frac{N_e \cos[ b\,\omega] + \cosh[a_A\;\omega]+e^{-a_B|\omega|}-e^{-(a_B-1)|\omega|} +\frac{1}{2}e^{-\frac{|\omega|}{2}}+\frac{1}{2}}{\sqrt{2\pi}\cosh[\frac{\omega}{2}]}.\eea
 
The solution for $d_A$ real can be obtained by analytical continuation of the above result, as mentioned above. The number of roots is given by formula same as \eqref{nrootssr}, from which we obtain the $z$-component of the spin of this state as \bea(S^z)_{us}=\frac{1}{2}.\eea 
  
This state has fermionic parity $\mathcal{P}=-1$.  The impurity at the left boundary is screened due to the Kondo effect while the impurity at the right boundary is unscreened. Due to the $SU(2)$ symmetry we immediately deduce that there is another ground state in the same fermion parity sector, degenerate with the above, which has the opposite spin \bea
(S^z)_{\widehat{usA}}=-\frac{1}{2}.
\eea

Actually, this state can be obtained by choosing the Bethe reference state with all spins down instead of up \cite{korepin1993quantum}. Hence there exists two degenerate states $\ket{\rm{K;\!US}}_{1/2}$, $\ket{\rm{K;\!US}}_{-1/2}$ in the K-YSR phase with spins $S^z=1/2$ and $S^z=-1/2$ respectively. They are both described by the distribution $\tilde\rho_\text{usA}(\omega)$. 

\smallskip

There exists another state $\ket{\rm{K;\!S}}$ in this phase. This state can be obtained by adding a boundary string to either of the states $\ket{\rm{S;\!US}}_{1/2}$, $\ket{\rm{K;\!US}}_{-1/2}$. The boundary strings arise as purely imaginary solutions of the Bethe equations. These purely imaginary Bethe roots, which correspond to the bound states, appear as poles in the dressed or physical boundary S-matrix  \cite{gosz,skorik}. We categorize the boundary strings as short boundary strings and wide boundary strings if the absolute value of the imaginary part is lesser or greater than one respectively. By observation we see that, in the limit $N_e\rightarrow\infty$, for $a_B>1/2$, the Bethe equations \eqref{BAE} have a {\it unique} solution 
\be
\lambda_{bsB}=\pm i(a_B-1/2),
\label{boundstring}
\ee 
as the two $\pm$ strings lead to the same state by reflection symmetry.  $|Im(\lambda_{bs})|<1$ for $1/2<a_B<3/2$, hence it is a short or close boundary string in the K-YSR phase.

\smallskip 

Adding the boundary string \eqref{boundstring} to the Bethe equations \eqref{BAE} for $1/2<a_B<3/2$ and taking the logarithm, we obtain 

\begin{flalign} \nonumber\sum_{\sigma=\pm} N_e\Theta(\lambda_\alpha+\sigma b,1/2)+\Theta(\lambda_\alpha,a_A+1/2)+\Theta(\lambda_\alpha,1/2-a_A)-\Theta(\lambda_\alpha,a_B-1/2)+\Theta(\lambda_\alpha,1)\\ \label{Logbae4}=\Theta(\lambda_\alpha,3/2-a_B)+\sum_{\beta=-M}^{M} \Theta\left(\lambda_\alpha- \lambda_\beta,1\right) +\pi I_j. \end{flalign}
 
The above equation can be solved by following the same procedure as above, we obtain
  \bea
 \tilde{\rho}^b_{bsB}(\omega)=\tilde{\rho}_{usB}(\omega)+ \Delta\tilde{\rho}^b_{bsB}(\omega), 
 \eea
where the shift $\Delta\tilde{\rho}^b_{bsB}(\omega)$ is due to the presence of the boundary string which is given by
\bea  \label{BetherootsBS} \Delta\tilde{\rho}^b_{bsB}(\omega)= \frac{e^{-(1-a_B)|\omega|}+e^{-a_B|\omega|}}{2\sqrt{2\pi}\cosh[\omega/2]}. \eea

In the presence of the boundary string, the relation between the number of Bethe roots and the density distribution also takes a different form as compared to \eqref{nrootssr}. Namely
\bea 
\label{nrootsbs}2M^b_{bsB}-1=\int_{-\infty}^{+\infty} d\lambda\; \rho^b_{bsB}(\lambda),
\eea
from which, using $(S^z)^b_{bsB}=N/2-M^b_{bsB}$, we find
\bea 
(S^z)^b_{bsB}=0. 
\eea 
Thus the resulting state $\ket{\rm{K;\!S}}$ described by the Bethe root distribution $\tilde{\rho}^b_{bsB}(\omega)$ which has a boundary bound state has total spin $\vec{S}_T=0$ and has odd fermionic parity $\mathcal{P}=-1$. In this state the left impurity is screened due to Kondo effect where as the right impurity is screened due to local spin accumulation.

To get the energy of this state, or of the boundary string, we notice that it is given by the energy difference, up to chemical potential, between  the ground states with $S^z=0$ and $S^z=\pm 1/2$
\begin{eqnarray}
\label{ebk}E_B=E^0_{N_e}-\frac{1}{2}(E^0_{N_e-1}+E^0_{N_e+1}).
\end{eqnarray}

Here $E^0_{N_e}$ refers to the energy of the state with even number of particles which, in our system, corresponds to the state $\ket{\rm{K;\!S}}$ which includes the boundary string and has spin $S^z=0$. Similarly $E^0_{N_e+1}$ and  $E^0_{N_e-1}$ refer to the energies of the states $\ket{0,1/2}$ or $\ket{0,-1/2}$ with odd number of particles and spin $S^z=\pm 1/2$. The expression \eqref{ebk} is defined in \cite{Keselman2015} as the binding energy, which precisely measures the energy cost of adding an electron to the system. To calculate $E_B$ we use \eqref{logEnergysup}, from which we obtain the following expression for total energy of a state with $N_e$ fermions
 \bea
 E=\sum_{j=1}^{N_e}\frac{\pi}{L}n_j+\frac{iD}{2}
 \int_{-\infty}^{\infty}d\lambda\; \Theta\left(b-\lambda,1/2\right)\rho(\lambda). \nonumber \\
 \label{toten}
\eea
  From  \eqref{ebk} we find that $E_B$ has two contributions, one from the charge degrees of freedom, 
 and one from the spin degrees of freedom: 
 $E_B=E_{\text{charge}}+\epsilon$. The charge contribution is given by the charging energy
\begin{eqnarray}
  E_{\text{charge}}=\sum_{j=1}^{N_e} \frac{\pi}{L}n_j-\frac{1}{2}\left(\sum_{j=1}^{N_e+1} \frac{\pi}{L}n_j+\sum_{j=1}^{N_e-1} \frac{\pi}{L}n_j\right). \nonumber \\
  \label{totenc}
 \end{eqnarray}
 
 Note that the the charge quantum numbers take all the values from the cutoff $-DL$ upwards. In the ground state they fill all the slots from $n_j=-N_e \; \text{to}\; n_j=-1$. In the state with one extra particle they fill all the slots from $n_j=-N_e \;\text{to}\; n_j=0$. In the state with one less particle there is an unfilled slot at $n_j=-1$ which corresponds to a holon excitation. We obtain
 \bea \label{bounden}E_{\text{charge}}=-\frac{\pi}{2L},\eea
 
 hence it vanishes in the thermodynamical limit. The spin contribution is given by the expression
 \bea
 \epsilon_B=E_{0}+\frac{iD}{2}\int_{-\infty}^{+\infty} d\lambda\; \Theta\left(b-\lambda,1/2\right)\Delta\rho^b_{bsB}(\lambda), \nonumber \\
 \label{totens}
\eea

where $E_{0}=D\Theta(b-i(a-1/2),1/2)+D\Theta(b+i(a-1/2),1/2)$ and $\Delta\rho^b_{bs}(\lambda)$ is the shift of the Bethe roots distribution due  to the boundary string which is given in (\ref{BetherootsBS}). Evaluating (\ref{totens}) we find that the energy of the boundary string takes the form

\bea E_B=E_{\text{charge}}+\epsilon_B. \eea

The charge part of the energy is \bea\label{bounden}E_{\text{charge}}=-\frac{\pi}{2L},\eea

 whereas, the spin part of the energy of the boundary string is

\bea \label{midgapen} \epsilon_B=-\Delta\sin(a_B\pi).\eea

Hence the energy of the boundary string or equivalently the energy difference between the states $\ket{\rm{K;\!S}}$ and $\ket{\rm{K;\!US}}_{\pm 1/2}$ is always less than the bulk gap $\Delta$. For $1/2<a_B<1$, $\epsilon_B<0$, hence the ground state is $\ket{\rm{K;\!S}}$ which contains a boundary bound state localized at the edge where the impurity lives and screens it. Removing this boundary bound state would unscreen the impurity, which costs energy less than the bulk gap yielding the states $\ket{\rm{K;\!US}}_{\pm 1/2}$. For $1<a_B<3/2$, $\epsilon_B>0$, hence the ground state comprises of the states $\ket{\rm{K;\!US}}_{\pm 1/2}$ and is two-fold degenerate. At $a_B=1$, $\epsilon_B=0$ and hence the ground state is three-fold degenerate where a level crossing occurs between the states $\ket{\rm{K;\!S}}$ and $\ket{\rm{K;\!US}}_{\pm 1/2}$. We also find that the energy difference between the ground state $\ket{\text{K,K}}$ in the K-K phase and the ground state $\ket{\rm{K;\!S}}$ in the K-YSR phase vanishes as one approaches the boundary between these phases at $a_B=1/2$. 

\subsection{Kondo-US phase}
The \textit{Kondo-unscreened} (K-US) phase occurs for $a_B>\frac{3}{2}$ and $d_A$ real or imaginary with $a_A<\frac{1}{2}$. In this phase the impurity at the left boundary is in Kondo phase whereas the impurity at the right boundary is in its unscreened phase.  

 By considering the state with all real Bethe roots $\lambda_\alpha$, one again obtains the states $\ket{\rm{K;\!US}}_{1/2}$ and $\ket{\rm{K;\!US}}_{-1/2}$ which are given by the distribution $\tilde{\rho}_{usB}(\omega)$. The boundary string solution $\lambda_{bsB}=i(a_B-1/2)$ still exists but adding this solution to the states $\ket{\rm{K;\!US}}_{1/2}$ or $\ket{\rm{K;\!US}}_{-1/2}$ is not possible unless one adds spinons in the bulk. This is due to the fact that $|Im(\lambda_{bsB})|>1$ for $a_B>3/2$, making it a wide boundary string \cite{PRA}.  The energy of the state corresponding to the addition of the wide boundary string goes above the mass gap due to the presence of the bulk hole and hence it does not correspond to a mid-gap state. Hence in the US-S phase the ground state comprises of two states  $\ket{\rm{K;\!US}}_{1/2}$ and $\ket{\rm{K;\!US}}_{-1/2}$ and is two-fold degenerate.

\vspace{4mm}

The results for the phases YSR-S, US-S can be obtained from the results of the phases S-YSR and S-US respectively by interchanging $a_A\leftrightarrow a_B$.

\subsection{YSR-YSR phase}

This phase occurs for $\frac{1}{2}<a_A,a_B<\frac{3}{2}$ and both impurities are in $YSR$ phase. There are four sub-phases depending on the values of $a_A,a_B$. Consider the state with all real $\lambda_\alpha$. Applying logarithm to Bethe equations \eqref{BAE} we obtain

\begin{flalign}\nonumber \sum_{\sigma=\pm} N_e\Theta(\lambda_\alpha+\sigma b,1/2)+\sum_{m=A,B}\Theta(\lambda_\alpha,a_m+1/2)-\Theta(\lambda_\alpha,a_m-1/2)\\+\Theta(\lambda_\alpha,1) 
=\sum_{\beta=-M}^{M} \Theta\left(\lambda_\alpha- \lambda_\beta,1\right) +\pi I_j. \label{Logbae4}\end{flalign}
 
 Differentiating \eqref{Logbae4} and following the same procedure as before we obtain

\begin{eqnarray}\label{gmdensity2}
g_{usAB}(\lambda)&=&\rho_\text{usAB}(\lambda)+\int_{-\infty}^{\infty}\mathrm{d}\mu\,\varphi(\lambda-\mu,1)\rho_\text{s}(\mu),
\end{eqnarray}

where $g_{usB}(\lambda)=\sum_{\sigma=\pm}N_e\varphi(\lambda+\sigma b ,1/2) +\sum_{m=A,B}\varphi(\lambda, a_m+1/2)-\varphi(\lambda, a_m-1/2)+\varphi(\lambda,1/2)+\varphi(\lambda,1)$.

 Solving \eqref{gmdensity2} by Fourier transform we obtain the following Fourier transformed density distribution of roots

\bea  \label{denscreened4}\tilde\rho_\text{usAB}(\omega)= \frac{N_e \cos[ b\,\omega] +\sum_{m=A,B}e^{-a_m|\omega|}-e^{-(a_m-1)|\omega|} +\frac{1}{2}e^{-\frac{|\omega|}{2}}+\frac{1}{2}}{\sqrt{2\pi}\cosh[\frac{\omega}{2}]}.\eea
 
The number of roots is given by the formula same as in \eqref{nrootssr}, from which we obtain the $z$-component of the spin of this state \bea(S^z)_{usAB}=1.\eea 
  
This state has fermionic parity $\mathcal{P}=1$.  Both the impurities are unscreened. Due to the $SU(2)$ symmetry, there exists another ground state in the same fermion parity sector, degenerate with the above, which has the opposite spin \bea
(S^z)_{\widehat{usA}}=-1.
\eea

This state can be obtained by choosing the Bethe reference state with all spins down instead of up. Hence there exist two degenerate states $\ket{\rm{US;\!US}}_{1/2,1/2}$, $\ket{\rm{US;\!US}}_{-1/2,-1/2}$ in the YSR-YSR phase with spins $1$ and $-1$ respectively, which are both described by the distribution $\tilde\rho_\text{usAB}(\omega)$. There exists two other states in which the spins of the two impurities are opposite, that is $\ket{US, US}_{1/2,-1/2}$ and $\ket{\rm{US;\!US}}_{-1/2,1/2}$. The construction of these states will be explained at the end of this section.

\smallskip

There exists a state $\ket{\rm{US;\!S}}_{1/2}$ in this phase. This state can be obtained by adding a boundary string $\lambda_{bsB}$ to the state $\ket{\rm{US;\!US}}_{1/2,1/2}$. 
Adding this to the Bethe equations \eqref{BAEsup} for $1/2<a_B<3/2$ and taking the logarithm we obtain 

\begin{flalign} \nonumber\sum_{\sigma=\pm} N_e\Theta(\lambda_\alpha+\sigma b,1/2)+\Theta(\lambda_\alpha,a_A+1/2)-\Theta(\lambda_\alpha,a_A-1/2)-\Theta(\lambda_\alpha,a_B-1/2)+\Theta(\lambda_\alpha,1)\\ \label{Logbae4}=\Theta(\lambda_\alpha,3/2-a_B)+\sum_{\beta=-M}^{M} \Theta\left(\lambda_\alpha- \lambda_\beta,1\right) +\pi I_j. \end{flalign}
 
The above equation can be solved by following the same procedure as above, we obtain
  \bea
 \tilde{\rho}^b_{bsB'}(\omega)=\tilde{\rho}_{usB'}(\omega)+ \Delta\tilde{\rho}^b_{bsB}(\omega), 
 \eea
where the shift $\Delta\tilde{\rho}^b_{bsB}(\omega)$ is due to the presence of the boundary string which is given by \ref{BetherootsBS}

%\bea  \label{BetherootsBS} \Delta\tilde{\rho}^b_{bsB}(\omega)= \frac{e^{-(1-a_B)|\omega|}+e^{-a_B|\omega|}}{2\sqrt{2\pi}\cosh[\omega/2]} \eea

In the presence of the boundary string, the relation between the number of Bethe roots and the density distribution also takes a different form as compared to \eqref{nrootssr}. This is given by
\bea 
\label{nrootsbs}2M^b_{bsB'}-1=\int_{-\infty}^{+\infty} d\lambda\; \rho^b_{bsB'}(\lambda),
\eea
from which, using $(S^z)^b_{bsB}=N/2-M^b_{bsB}$, we find
\bea 
(S^z)^b_{bsB'}=\frac{1}{2}. 
\eea 
Thus the resulting state $\ket{\rm{US;\!S}}_{1/2}$ described by the Bethe root distribution $\tilde{\rho}^b_{bsB'}(\omega)$  has total spin $\vec{S}_T=\frac{1}{2}$ and has odd fermionic parity $\mathcal{P}=-1$. There exists another degenerate state $\ket{\rm{US;\!S}}_{-1/2}$, that can be obtained by adding $\lambda_{bsB}$ to the state $\ket{\rm{US;\!US}}_{-1/2,-1/2}$. In both of the above states, the impurity at the left boundary is unscreened while the impurity at the right boundary is screened by a local spin accumulation. Similarly, there exists degenerate states $\ket{\rm{S;\!US}}_{1/2}$ and $\ket{\rm{S;\!US}}_{-1/2}$ that can be obtained by adding the boundary string $\lambda_{bsA}=\pm i(a_A-1/2)$ to the states $\ket{\rm{US;\!US}}_{1/2,1/2}$ and $\ket{\rm{US;\!US}}_{-1/2,-1/2}$ respectively. In these two states, the impurity at the left boundary is screened by a local spin accumulation while the impurity at the right boundary is unscreened.

\vspace{3mm}

There also exists a state $\ket{\rm{S;\!S}}$ which can be obtained by adding both the boundary strings $\lambda_{bsA}$ and $\lambda_{bsB}$ to either of the states 
$\ket{\rm{US;\!US}}_{1/2,1/2}$, $\ket{\rm{US;\!US}}_{-1/2,-1/2}$. The resulting state is described by the following density distribution

\bea
 \tilde{\rho}^b_{bsAB}(\omega)=\tilde{\rho}_{usAB}(\omega)+ \Delta\tilde{\rho}^b_{bsA}+\Delta\tilde{\rho}^b_{bsB}(\omega).
 \eea

In the presence of the two boundary strings, the relation between the number of Bethe roots and the density distribution takes a different form 
\bea 
\label{nrootsbs2}2M^b_{bsAB}-3=\int_{-\infty}^{+\infty} d\lambda\; \rho^b_{bsAB}(\lambda),
\eea
from which, using $(S^z)^b_{bsB}=N/2-M^b_{bsAB}$, we find
\bea 
(S^z)^b_{bsAB}=0. 
\eea 

Both impurities are screened by the boundary bound states localized at the respective impurities.

\vspace{3mm}

Consider the state $\ket{\rm{US;\!S}}_{1/2}$, which is obtained by adding the boundary string $\lambda_{bsB}$ to the state $\ket{\rm{US;\!US}}_{1/2,1/2}$. In the presence of this boundary string, there exists another boundary string solution $\lambda_{bsB'}=\pm i(\frac{3}{2}-a_B)$. We can add this boundary string solution to the state $\ket{\rm{US;\!S}}_{1/2}$. We obtain the state $\ket{\rm{US;\!US}}_{1/2,-1/2}$ which is described by the following density distribution 

\bea
 \tilde{\rho}^b_{bsB''}(\omega)=\tilde{\rho}_{bsB'}(\omega)+ \Delta\tilde{\rho}^b_{bsB'}(\omega).
 \eea

Here

\bea \Delta\tilde{\rho}^b_{bsB'}(\omega)= \frac{e^{-(2-a_B)|\omega|}+e^{-(a_B-1)|\omega|}}{2\sqrt{2\pi}\cosh[\omega/2]}. \eea

The number of roots is given by

\bea 
2M^b_{bsB''}-3=\int_{-\infty}^{+\infty} d\lambda\; \rho^b_{bsB''}(\lambda),
\eea
from which, using $(S^z)^b_{bsB''}=N/2-M^b_{bsB''}$, we find
\bea 
(S^z)^b_{bsAB}=0. 
\eea 

Similarly the state $\ket{\rm{US;\!US}}_{-1/2,1/2}$ can be obtained from the state $\ket{\rm{US;\!S}}_{-1/2}$ by following the same procedure described above. The energy of the boundary string $\lambda_{bsB'}$ is exactly equal and opposite in sign to that of $\lambda_{bsB}$. Hence, the states $\ket{\rm{US;\!US}}_{-1/2,1/2}$, $\ket{\rm{US;\!US}}_{1/2,-1/2}$ are degenerate with the states $\ket{\rm{US;\!US}}_{1/2,1/2}$, $\ket{\rm{US;\!US}}_{1/2,-1/2}$.

\vspace{4mm}

In the phase US-YSR, the left boundary string becomes a wide boundary string as $a_A>\frac{3}{2}$, and hence the left impurity cannot be screened. Only low lying states that exist are $\ket{\rm{US;\!US}}_{\pm 1/2,\pm 1/2}$ in which both impurities are unscreened and the states $\ket{\rm{US;\!S}}_{\pm 1/2}$, in which the right boundary is screened by the boundary bound state. The results in the phase YSR-US can be obtained from the results in phase US-YSR by the exchange $a_A\leftrightarrow a_B$. In the phase US-US, both impurities will remain unscreened and cannot be screened. The states $\ket{\rm{US;\!US}}_{1/2,-1/2}$ and $\ket{\rm{US;\!US}}_{-1/2,1/2}$ are obtained by adding the boundary strings $\lambda_{bsA}$,$\lambda_{bsB}$ which are now wide boundary strings in US-US phase as as $a_A,a_B>\frac{3}{2}$. The spin part of the energy of the wide boundary strings is exactly zero and hence the ground state is four fold degenerate in US-US phase.

\section{Effective algebra in YSR-YSR phase}

From the Bethe solution we have nine states in the phase where both the impurities are in their respective YSR phases. When the RG invariant parameters $a_A,a_B$ corresponding to both the impurities are equal to one, we have a nine fold degenerate ground state. In this phase, we can also specify the local spin quantum numbers. This is possible because we know whether each impurity is screened or unscreened and also the spin of the unscreened impurity based on the boundary Bethe roots describing each state. The existence of local spin quantum numbers gives rise to a rich boundary spectrum. 
 
\vspace{2mm}

1) There are four states in which both the impurities are unscreened. 

2) There are four states in which one impurity is screened and the other is unscreened.

3) There is one state in which both the impurities are screened.

\vspace{4mm}

 The algebra corresponding to these states is $SPL(2,1)$. In the following we briefly describe this algebra and construct the irreducible representation of the nine degenerate states.
 
 \vspace{2mm}
 
  There exists a baryon number $b$ and total spin quantum number $q$ which can be used to specify a representation $[b,q]$. Each representation $[b,q]$ may contain four multiplets  $|b,q,q_3\rangle, \;\; |b+\frac{1}{2},q-\frac{1}{2},q_3\rangle, \;\; |b-\frac{1}{2},q-\frac{1}{2},q_3\rangle, \;\; |b,q-1,q_3\rangle$, where $q_3=-q...q$, is the spin $S^z$. Different representations corresponding to different values of $b,q$ may or may not have certain multiplets \cite{Rittenberg} (We use the same notation as theirs).

There exists two odd operators $V_{\pm}, W_{\pm}$, and three even operators $Q_{\pm}, Q_3$, and $B$. The definition of odd and even operators is provided below (\ref{evenodd}). All the states are eigenstates of the Baryon number operator $B$, with eigenvalues being the respective baryon numbers. The operator $V_{\pm}$ increases the baryon number by $\frac{1}{2}$ and simultaneously increases/decreases the spin $S^z$ by $\frac{1}{2}$. The operator $W_{\pm}$ decreases the baryon number by $\frac{1}{2}$ and simultaneously increase/decrease the spin $S^z$ by $\frac{1}{2}$. The operator $Q_{\pm}$ does not change the baryon number but only increases/decreases the spin $S^z$ by $1$ and $Q_3$ is just measures the $S^z$.  These operators satisfy the following commutation relations

\begin{align} 
 [Q_3,Q_{\pm}]=\pm &Q_{\pm},\;[Q_{\pm},Q_{-}]=2Q_{3},\\
[B,Q_{\pm}]&=[B,Q_{3}]=0,\\
[B,V_{\pm}]=V_{\pm}/2,&\; [B,W_{\pm}]=-W_{\pm}/2,\\
[Q_3,V_{\pm}]=\pm V_{\pm}/2,&\; [Q_3,W_{\pm}]=\pm W_{\pm}/2,\\
[Q_{\pm},V_{\mp}]=V_{\pm}&,\; [Q_{\pm},W_{\mp}]=W_{\pm},
\end{align}
and

\begin{align}
[Q_{\pm},V_{\pm}&]=[Q_{\pm},W_{\pm}]=0,\\
\{V_{\pm},V_{\pm}\}&=\{V_{\pm},V_{\mp}\}=0,\\
\{W_{\pm},W_{\pm}\}=&\{W{\pm},W_{\mp}\}=0,\\
\{V_{\pm},W_{\pm}\}=\pm Q_{\pm}, \{&V_{\pm},W_{\mp}\}=-Q_3\pm B.
\end{align}

\vspace{3mm}
Using Wigner-Eckart theorem, one obtains
 \begin{align}
& V_{\pm}\ket{b,q,q_3}=\pm \alpha \sqrt{q\mp q_3}\ket{b+\frac{1}{2},q-\frac{1}{2},q_3\pm \frac{1}{2}},\\
& W_{\pm}\ket{b,q,q_3}=\pm \beta \sqrt{q\mp q_3}\ket{b-\frac{1}{2},q-\frac{1}{2},q_3\pm \frac{1}{2}},\\
& V_{\pm}\ket{b+\frac{1}{2},q-\frac{1}{2},q_3}=0,\\
& W_{\pm}\ket{b+\frac{1}{2},q-\frac{1}{2},q_3}=\gamma \sqrt{q\pm q_3+\frac{1}{2}}\ket{b,q,q_3\pm \frac{1}{2}}\pm \delta \sqrt{q\mp q_3-\frac{1}{2}}\ket{b,q-1,q_3\pm\frac{1}{2}},\\
& V_{\pm}\ket{b-\frac{1}{2},q-\frac{1}{2},q_3}=\epsilon \sqrt{q\pm q_3+\frac{1}{2}}\ket{b,q,q_3\pm \frac{1}{2}}\pm \xi\sqrt{q\mp q_3-\frac{1}{2}}\ket{b,q-1,q_3\pm\frac{1}{2}},\\
& W_{\pm}\ket{b-\frac{1}{2},q-\frac{1}{2},q_3}=0,\\
& V_{\pm}\ket{b,q-1,q_3}=\tau \sqrt{q\pm q_3}\ket{b+\frac{1}{2},q-\frac{1}{2},q_3\pm \frac{1}{2}},\\
& W_{\pm}\ket{b,q-1,q_3}=\omega \sqrt{q\mp q_3}\ket{b-\frac{1}{2},q-\frac{1}{2},q_3\pm \frac{1}{2}}.\label{Wignereckart}
\end{align}

Here $\alpha, \beta,...,\omega$ are some numbers which are independent of $q_3$ and are constrained and take certain values in any given representation $[b,q]$. At each edge, the total spin can be either $S^z=\pm \frac{1}{2}$ or $S=0$. The representation is $[\frac{1}{2},\frac{1}{2}]$. It can only contain the multiplets $|\frac{1}{2},\frac{1}{2},m\rangle, |1,0,0\rangle$. Using (\ref{Wignereckart}), we obtain the following relations 

\begin{equation}
 Q_+|\frac{1}{2},\frac{1}{2}, -\frac{1}{2}\rangle=|\frac{1}{2},\frac{1}{2}, \frac{1}{2}\rangle, \hspace{2mm} 
 Q_-|\frac{1}{2},\frac{1}{2}, \frac{1}{2}\rangle=|\frac{1}{2},\frac{1}{2}, -\frac{1}{2}\rangle, \end{equation}
and 
\bea\nonumber
V_{+} |\frac{1}{2},\frac{1}{2}, -\frac{1}{2}\rangle=\alpha_{\frac{1}{2}}|1,0,0\rangle, \hspace{2mm} V_{-} |\frac{1}{2},\frac{1}{2}, \frac{1}{2}\rangle=-\alpha_{\frac{1}{2}}|1,0,0\rangle,\\ W_{+} |1,0,0\rangle= \gamma_{\frac{1}{2}} |\frac{1}{2},\frac{1}{2}, \frac{1}{2}\rangle \hspace{2mm} W_{-} |1,0,0\rangle=\gamma_{\frac{1}{2}} |\frac{1}{2},\frac{1}{2},- \frac{1}{2}\rangle, \eea

with the constraint $\alpha_{\frac{1}{2}}\gamma_{\frac{1}{2}}=1$. The operators acting on the other states yield zeros. 

\hspace{10mm}

The irreducible representation of $[\frac{1}{2},\frac{1}{2}]\otimes [\frac{1}{2},\frac{1}{2}]$ is $[1,1] \oplus [\frac{3}{2},\frac{1}{2}]$.  $[1,1]$ contains the multiplets $|1,1,m\rangle, |\frac{3}{2},\frac{1}{2},m\rangle$. $[\frac{3}{2},\frac{1}{2}]$ contains the multiplets $|\frac{3}{2},\frac{1}{2},m\rangle, |2,0,0\rangle, |1,0,0\rangle$.

\vspace{5mm}

Using (\ref{Wignereckart}) we obtain the following relations for $[1,1]$
\bea\nonumber
Q_{+} |1,1,0\rangle=\sqrt{2} |1,1,1\rangle, \hspace{2mm} Q_{+} |1,1,-1\rangle= \sqrt{2} |1,1,0\rangle, \\ Q_{-} |1,1,1\rangle=\sqrt{2} |1,1,0\rangle, \hspace{2mm} Q_{-} |1,1,0\rangle=\sqrt{2} |1,1,-1\rangle, \eea

\bea
Q_+|\frac{3}{2},\frac{1}{2}, -\frac{1}{2}\rangle=|\frac{3}{2},\frac{1}{2},\frac{1}{2}\rangle, \hspace{3mm} Q_-|\frac{3}{2},\frac{1}{2}, \frac{1}{2}\rangle=|\frac{3}{2},\frac{1}{2},-\frac{1}{2}\rangle,\eea

and 

\bea
V_{-}|1,1,1\rangle=-\alpha_{1}\sqrt{2}|\frac{3}{2},\frac{1}{2},\frac{1}{2}\rangle, \hspace{2mm} V_{-}|1,1,0\rangle=-\alpha_{1}|\frac{3}{2},\frac{1}{2},-\frac{1}{2}\rangle, \\ V_{+}|1,1,0\rangle=\alpha_{1} |\frac{3}{2},\frac{1}{2},\frac{1}{2}\rangle, \hspace{2mm} V_{+}|1,1,-1\rangle= \alpha_{1}\sqrt{2} |\frac{3}{2},\frac{1}{2},-\frac{1}{2}\rangle,\\
W_{+}|\frac{3}{2},\frac{1}{2},\frac{1}{2}\rangle=\gamma_{1}\sqrt{2} |1,1,1\rangle, \hspace{2mm} W_{+}|\frac{3}{2},\frac{1}{2},-\frac{1}{2}\rangle=\gamma_{1} |1,1,0\rangle, \\ W_{-}|\frac{3}{2},\frac{1}{2},\frac{1}{2}\rangle=\gamma_{1} |1,1,0\rangle, \hspace{2mm} W_{-}|\frac{3}{2},\frac{1}{2},-\frac{1}{2}\rangle=\gamma_{1}\sqrt{2} |1,1,-1\rangle, \eea

with the constraint $\alpha_{1}\gamma_{1}=1$. The operators acting on the other states yield zeros. 

\vspace{3mm}

Similarly for $[\frac{3}{2},\frac{1}{2}]$, we obtain,
\begin{equation}
Q_{+}|\frac{3}{2},\frac{1}{2},-\frac{1}{2}\rangle =|\frac{3}{2},\frac{1}{2},\frac{1}{2}\rangle, \hspace{2mm} Q_{-}|\frac{3}{2},\frac{1}{2},\frac{1}{2}\rangle =|\frac{3}{2},\frac{1}{2},-\frac{1}{2}\rangle, \end{equation}
and

\bea\nonumber
V_{+}|\frac{3}{2},\frac{1}{2},-\frac{1}{2}\rangle = \alpha_{\frac{3}{2}}|2,0,0\rangle, \hspace{2mm} V_{-}|\frac{3}{2},\frac{1}{2},\frac{1}{2}\rangle = -\alpha_{\frac{3}{2}}|2,0,0\rangle ,\\ W_{+}|\frac{3}{2},\frac{1}{2},-\frac{1}{2}\rangle = \beta_{\frac{3}{2}}|1,0,0\rangle, \hspace{2mm} W_{-}|\frac{3}{2},\frac{1}{2},\frac{1}{2}\rangle = -\beta_{\frac{3}{2}}|1,0,0\rangle,\\
W_{+} |2,0,0\rangle =\gamma_{\frac{3}{2}}  |\frac{3}{2},\frac{1}{2},\frac{1}{2}\rangle, \hspace{2mm} W_{-} |2,0,0\rangle =\gamma_{\frac{3}{2}}  |\frac{3}{2},\frac{1}{2},-\frac{1}{2}\rangle,\\ V_{+} |1,0,0\rangle =\epsilon_{\frac{3}{2}}  |\frac{3}{2},\frac{1}{2},\frac{1}{2}\rangle,  \hspace{2mm} V_{-} |1,0,0\rangle =\epsilon_{\frac{3}{2}}  |\frac{3}{2},\frac{1}{2},-\frac{1}{2}\rangle, \eea

with the constraint $\alpha_{\frac{3}{2}}\gamma_{\frac{3}{2}}=2$ and  $\beta_{\frac{3}{2}}\epsilon_{\frac{3}{2}}=-1$, and the operators acting on the other states yielding zeros. 

\hspace{5mm}

When an operator $M$ acts on the state $|f\rangle \otimes |g\rangle$ one obtains 

\begin{align}  &M(|f\rangle \otimes |g\rangle) =(M|f\rangle)\otimes |g\rangle \pm \; |f\rangle \otimes (M|g\rangle \label{evenodd}\end{align}

If the operator $M$ is even then we only have a $+$ sign. If $M$ is odd, then the sign depends on the evenness or oddness of the state $|f\rangle$. If the state with baryon number $b$ is chosen to be even (odd) then the state with baryon number $b+\frac{1}{2}$ is odd (even). We choose $|\frac{1}{2},\frac{1}{2},m\rangle$ to be even. 

\hspace{4mm}

Consider $[1,1]$. Here the multiplet $|1,1,m\rangle$ must be describing the triplet states where the two impurities are unscreened. Let us now assume a certain form for the multiplet $|\frac{3}{2},\frac{1}{2},m\rangle$ in terms of the product states and see if we obtain the triplet states $|1,1,m\rangle$, by operating with $V_{\pm}, W_{\pm}$. Let

\begin{equation}
 |\frac{3}{2},\frac{1}{2},\pm\frac{1}{2}\rangle_+=\frac{1}{\sqrt{2}}\left(|1,0,0\rangle \otimes |\frac{1}{2},\frac{1}{2},\pm\frac{1}{2}\rangle + |\frac{1}{2},\frac{1}{2},\pm\frac{1}{2}\rangle \otimes |1,0,0\rangle\right). \end{equation}

Here we used the subscript $+$ to denote the symmetric superposition. By acting with $W_{+}$ on $|\frac{3}{2},\frac{1}{2},\frac{1}{2}\rangle$ and using the relations (2) and (6), we obtain 
\begin{equation}
|1,1,1\rangle = \left(\frac{\gamma_{\frac{1}{2}}}{\gamma_{1}}\right)|\frac{1}{2},\frac{1}{2},\frac{1}{2}\rangle \otimes |\frac{1}{2},\frac{1}{2},\frac{1}{2}\rangle.\end{equation}

Similarly acting with $W_{+}$ on $|\frac{3}{2},\frac{1}{2},-\frac{1}{2}\rangle$ we obtain 

\begin{equation}
|1,1,0\rangle =\frac{1}{\sqrt{2}}\left(\frac{\gamma_{\frac{1}{2}}}{\gamma_{1}}\right)\left(|\frac{1}{2},\frac{1}{2},\frac{1}{2}\rangle \otimes |\frac{1}{2},\frac{1}{2},-\frac{1}{2}\rangle+|\frac{1}{2},\frac{1}{2},-\frac{1}{2}\rangle \otimes |\frac{1}{2},\frac{1}{2},\frac{1}{2}\rangle \right).\end{equation}

Acting with $W_{-}$ on $|\frac{3}{2},\frac{1}{2},-\frac{1}{2}\rangle$ we obtain 
\begin{equation}
|1,1,-1\rangle =\left(\frac{\gamma_{\frac{1}{2}}}{\gamma_{1}}\right)|\frac{1}{2},\frac{1}{2},-\frac{1}{2}\rangle \otimes |\frac{1}{2},\frac{1}{2},-\frac{1}{2}\rangle.\end{equation}

Now by acting with $V_{-}$ on the state $|1,1,1\rangle$, we obtain

\begin{equation}
|\frac{3}{2},\frac{1}{2},\frac{1}{2}\rangle_{+}=\frac{1}{\sqrt{2}}\left(\frac{\alpha_{\frac{1}{2}}}{\alpha_{1}}\right)\left(\frac{\gamma_{\frac{1}{2}}}{\gamma_{1}}\right)\left(|1,0,0\rangle \otimes |\frac{1}{2},\frac{1}{2},\frac{1}{2}\rangle + |\frac{1}{2},\frac{1}{2},\frac{1}{2}\rangle \otimes |1,0,0\rangle\right). \end{equation}

By acting with $V_{+}$ on the state $|1,1,-1\rangle$, we obtain

\begin{equation}
|\frac{3}{2},\frac{1}{2},-\frac{1}{2}\rangle_{+}=\left(\frac{\alpha_{\frac{1}{2}}}{\alpha_{1}}\right)\left(\frac{\gamma_{\frac{1}{2}}}{\gamma_{1}}\right)\left(|1,0,0\rangle \otimes |\frac{1}{2},\frac{1}{2},-\frac{1}{2}\rangle + |\frac{1}{2},\frac{1}{2},-\frac{1}{2}\rangle \otimes |1,0,0\rangle\right).\end{equation}

Using the constraints $\alpha_{1}\gamma_{1}=1$ and $\alpha_{\frac{1}{2}}\gamma_{\frac{1}{2}}=1$, we see that the above relations are self consistent. We can check  by repeatedly applying all the operators that all the relations are consistent. We can fix $\alpha_{\frac{1}{2}}=\alpha_{1}= \gamma_{\frac{1}{2}}=\gamma_{1}=1$.

\vspace{10mm}

In the representation $[3/2,1/2]$, we have $|1,0,0\rangle$. It is natural that this is a singlet of the two unscreened impurities. Let us check if we obtain this starting from the assumption that 

\begin{equation}
|\frac{3}{2},\frac{1}{2},\pm\frac{1}{2}\rangle_{-}=\frac{1}{\sqrt{2}}\left(|1,0,0\rangle \otimes |\frac{1}{2},\frac{1}{2},\pm\frac{1}{2}\rangle - |\frac{1}{2},\frac{1}{2},\pm\frac{1}{2}\rangle \otimes |1,0,0\rangle\right). \end{equation}

Here the subscript $-$ denotes the anti-symmetric superposition. Applying $W_{-}$ to this state we obtain 
\begin{equation}
|1,0,0\rangle =\frac{1}{\sqrt{2}}\left(\frac{\gamma_{\frac{1}{2}}}{\beta_{\frac{3}{2}}}\right)\left(|\frac{1}{2},\frac{1}{2},\frac{1}{2}\rangle \otimes |\frac{1}{2},\frac{1}{2},-\frac{1}{2}\rangle-|\frac{1}{2},\frac{1}{2},-\frac{1}{2}\rangle \otimes |\frac{1}{2},\frac{1}{2},\frac{1}{2}\rangle \right).\end{equation}

By applying $V_{+}$ to this state we obtain 

\begin{equation}|\frac{3}{2},\frac{1}{2},\frac{1}{2}\rangle_{-}=-\frac{1}{\sqrt{2}}\left(\frac{\gamma_{\frac{1}{2}}}{\beta_{\frac{3}{2}}}\right) \left(\frac{\alpha_{\frac{1}{2}}}{\epsilon_{\frac{3}{2}}}\right)\left(|1,0,0\rangle \otimes |\frac{1}{2},\frac{1}{2},\frac{1}{2}\rangle - |\frac{1}{2},\frac{1}{2},\frac{1}{2}\rangle \otimes |1,0,0\rangle\right) \end{equation}

Hence with the  constraint $\beta_{\frac{3}{2}}\epsilon_{\frac{3}{2}}=-1$ and $\alpha_{\frac{1}{2}}\gamma_{\frac{1}{2}}=1$, we see that we have consistency. We make the choice $\alpha_{\frac{1}{2}}=\alpha_{1}= \gamma_{\frac{1}{2}}=\beta_{\frac{3}{2}}=1$ and $\epsilon_{\frac{3}{2}}=-1$. By applying $V_{+}$ to $|\frac{3}{2},\frac{1}{2},\frac{1}{2}\rangle$ we obtain

\begin{equation}
|2,0,0\rangle = -\sqrt{2} \left(\frac{\alpha_{\frac{1}{2}}}{\alpha_{\frac{3}{2}}}\right) \left(|1,0,0\rangle\otimes |1,0,0\rangle\right).\end{equation}

By applying $W_{\pm}$ to this state we obtain 

\begin{equation}
|\frac{3}{2},\frac{1}{2},\pm\frac{1}{2}\rangle_{-}=\sqrt{2}\left(\frac{\gamma_{\frac{1}{2}}}{\gamma_{\frac{3}{2}}}\right) \left(\frac{\alpha_{\frac{1}{2}}}{\alpha_{\frac{3}{2}}}\right)\left(|1,0,0\rangle \otimes |\frac{1}{2},\frac{1}{2},\pm\frac{1}{2}\rangle - |\frac{1}{2},\frac{1}{2},\pm\frac{1}{2}\rangle \otimes |1,0,0\rangle\right). \end{equation}

Hence we see that with the choice $\alpha_{\frac{3}{2}}=-\sqrt{2}, \gamma_{\frac{3}{2}}=-\sqrt{2}$, we have consistency.

\end{widetext}
\end{document}